\documentstyle[epsf,12pt]{article}

\newcommand{\grs}{$\gamma$-rays }
\newcommand{\gr}{$\gamma$-ray }

\newcommand{\alx}{\alpha_{\rm x} }
\newcommand{\ale}{\alpha_{\rm e} }

\newcommand{\alr}{\alpha_{\rm r} }
\newcommand{\alinj}{\alpha_{\rm inj} }

\begin{document}

\centerline{\Large Modelling of the nonthermal flares in the galactic}
\centerline{\Large microquasar GRS 1915+105}

\vspace{3mm}
 
\centerline{\large A. M.~Atoyan$^{1,2}$ and F. A.~Aharonian$^1$} 

\vspace{2mm}

\centerline{ $^1$ Max-Planck-Institut f\"{u}r Kernphysik, Postfach 103980,
 D-69029 Heidelberg, Germany}
\centerline{ $^2$ Yerevan Physics Institute, Alikhanian Brothers 2,
375036 Yerevan, Armenia}

\vspace{10mm}

\centerline{ {\large {\bf Abstract}}}

\vspace{4mm}

\noindent
The microquasars recently discovered in our Galaxy offer unique opportunity for
a deep insight into the physical processes in relativistic jets observed in
different source populations. We 
study the temporal and spectral evolution of the  radio flares detected from
the relativistic ejecta in the microquasar GRS 1915+105, and propose a model 
which suggests that these flares are  
due to synchrotron radiation of relativistic electrons
suffering adiabatic and escape  losses 
in fastly expanding plasmoids (the radio clouds).
Analytical solutions to the 
kinetic equation for relativistic electrons in the expanding magnetized
clouds are found, and the synchrotron radiation of these electrons is
calculated. Detailed comparison of the calculated radio fluxes with the ones 
detected from the prominent flare of GRS 1915+105 during March/April 1994
provides conclusive information on the basic parameters in the ejecta, 
such as the absolute values  and temporal evolution of
the magnetic field, speed of expansion, the
rate of continuous injection of relativistic electrons into and their
energy-dependent escape from the clouds, etc.
The data of radio monitoring of the pair of resolved ejecta enable unambiguous
determination of  parameters of the bulk motion of counter ejecta and  the
degree of asymmetry between them, as well as contain important information
on the prime energy source for accelerated electrons, in particular, may
distinguish between the scenarios of bow shock powered and relativistic
magnetized wind powered plasmoids.  Assuming that the electrons in the ejecta
might be accelerated up to very high energies,
we calculate the fluxes of synchrotron radiation up to hard X-rays/soft
$\gamma$-rays,
 and inverse Compton GeV/TeV $\gamma$-rays expected during radio
flares, and discuss the implications which could follow from
either positive detection or flux upper limits of these energetic photons.

\vspace{5mm}


\vspace{3cm}

{\small
\noindent
e-mails: \\
A.Atoyan -- atoyan@fel.mpi-hd.mpg.de\\
F.Aharonian -- aharon@fel.mpi-hd.mpg.de}

\newpage

\section{Introduction}

The hard X-ray transient GRS~1915+105 discovered in 1992 
(Castro-Tirado et al. 1992) belongs to the class of  
galactic black hole (BH)
candidate sources characterized by an estimated mass of the compact 
object $M \geq 3 M_\odot$, strong variability of the electromagnetic radiation,
with the spectra extending to energies beyond $100 \, \rm keV$
(e.g. see Harmon et al. 1994). 
The invoking similarity of  these sources with AGNs, the objects 
powered by the accreting BHs of essentially larger mass scales,
has recently found a new convincing evidence, after the discovery of apparent
superluminal jets in GRS~1915+105 (Mirabel \& Rodriguez 1994, 
hereafter MR94),
and in another hard X-ray transient,  GRO J1655-40
(Tingay et al. 1995, Hjellming \& Rupen 1995). The existence of 
relativistic jets in these galactic sources, called ``microquasars'', 
makes them unique cosmic laboratories for a deeper insight into the  
complex phenomenon of the jets common in those powerful extragalactic objects 
(Mirabel \& Rodriguez 1995).  Indeed, being much closer
to us than AGNs, the microquasars offer an opportunity for monitoring of 
the jets in a much shorter spatial and temporal scales. 
Furthermore, the parameter $L_{\rm Edd}/d^2$,
where t$L_{\rm Edd} \simeq 1.3 \times 10^{38} (M/M_\odot) \, \rm erg/s\,$ 
is the Eddington luminosity (an indicator of 
the potential power of the source),     
and $d$ is the distance to the source, is typically by 2-4 orders
of magnitude higher for the galactic BH candidates than for AGNs. 
This enables 
detection of the galactic superluminal jets with  
Doppler factors even as small as $\delta \leq 1$,
in contrast to the case of jets in blazars 
(AGNs with jets close to the line of sight), for detection of which strong
Doppler boosting of the radiation, due to $\delta \gg 1$, is generally needed. 
Importantly, for the jets with aspect angles close 
to $90^\circ$ (which is the case in both GRS 1915+105 and GRO J1655-40), 
the Doppler factors of the approaching and receding components are comparable,
so both these components may be detectable 
(see Mirabel \& Rodriguez 1995).
 
In this paper we present the results of theoretical study of the
nonthermal flares of GRS 1915+105, which are observed in the
radio (MR94; Mirabel \& Rodriguez 1995; Rodriguez et al. 1995; 
Foster et al. 1996) and possibly also infra-red 
(Sams et al. 1996; Fender et al. 1997) bands, and can be expected  
at higher photon energies as well, if the electrons in the ejecta 
are accelerated up to very high 
energies (VHE). 
The main part of the paper is devoted to the explanation of the 
 radio flares observed, which provide 
important information about the principal parameters and processes in the 
radio emitting plasmoids (i.e. the clouds containing relativistic electrons 
and magnetic fields). In Section 2 we deduce, on the basis of 
qualitative consideration, the principal scenario for interpretation
of these flares. In Section 3 we obtain analytical solutions to the 
kinetic equation which
describes the temporal evolution of relativistic 
electrons in an expanding magnetized medium. 
In Section 4 we carry out detailed comparison of the fluxes of 
synchrotron radiation emerging from the model radio clouds
with the temporal
and spectral characteristics of the fluxes observed during the prominent
 March/April  1994 flare of GRS 1915+105 (MR94), which provides  
conclusive constraints for the model parameters of the  ejecta.
The data of radio monitoring of the two (approaching and receding) ejecta 
present particular interest. We show that these data not
only enable unambiguous determination of the aspect angle and the speed of
propagation of the twin ejecta (which perhaps are a little asymmetrical,
as follows from the observed flux ratio of the counter components), 
but also contain an important information
about the prime energy source for continuous energization of the plasmoids. 

For any reasonable assumption about the magnetic field in the ejecta, 
the energy spectrum of relativistic electrons should extend 
at least up to energies of a few GeV. It is not excluded, however, that 
the electrons in the ejecta are accelerated beyond these energies.
Therefore, after determination of the range of basic parameters of the 
expanding plasmoids, in Section 5  we speculate that in 
GRS 1915+105, similar to the case of
BL Lacs (e.g. Urry \& Padovani 1995; Ghisellini \& Maraschi 1997),
the electrons may be accelerated beyond TeV energies, 
$E\geq 10^{12}\,\rm eV$, and discuss the 
fluxes of synchrotron radiation which could then extend 
up to X-ray/\gr energies, as well as the fluxes of the inverse Compton (IC) 
$\gamma$-rays expected in the high and very high energies.
In Section 6 we summarize the main results of the paper.

\section{Basic estimates}

\subsection{Radio fluxes during outbursts}

Observations of GRS 1915+105 show that the radio emission of the 
source in its active state 
is characterized by the 
increase of the radio emission at $\nu \sim (1-10)\,\rm GHz$ from the flux 
level $S_{\nu} \leq 10 \,\rm mJy$ to $S_{\nu} \sim 100 \,\rm mJy$  
(the {\it plateau}
state, according to Foster et al. 1996), on top of which the episodes of radio 
{\it flares}, with durations from days to a month and fluxes 
up to $\sim 1\,\rm Jy\,$, are superimposed. 
 The spectra in the plateau state are 
flat or inverted, $\alr \leq 0$ (approximating $S_{\nu} \propto \nu^{-\alr}$), 
which is a typical indicative for a strong synchrotron self-absorption
in the source (e.g. Pacholczyk 1970). 
The radio flares, which most probably are related with the events of 
ejection of radio clouds, are distinguished 
by very rapid, less than a day, rise of the fluxes by several times,  
which then decay on timescales of days or more. Remarkably, already at  
the rising stage of the flares the spectra reveal transition
from optically thick ($\alr \leq 0$) to optically thin ($\alr > 0$) emission.
Initially, when the fluxes are at the maximum level,  
the spectra are typically hard, with $\alr \sim 0.5$, which may then  
steepen to $\alr \sim 1$. 

Convincing examples of such spectral behavior 
are given by Rodriguez et al (1995) where the detailed data of 
radio monitoring of 
GRS 1915+105 at different wavelengths during 5 months 
from 1 December 1993 to 30 April 1994 are summarized. 
Thus,  from Table 1 of Rodriguez et al. (1995) it is seen that 
the spectrum of the flare on December 6.0, 1993,
measured with VLA radio telescopes {\it simultaneously} at 5 different 
frequencies from $\approx 1.47\,\rm GHz$ to 22.49\,GHz was self-absorbed in 
the range of $1.47 \leq \nu \leq 4.89 \,\rm GHz$ ($\alr \approx -0.73$), 
whereas at higher 
frequencies it corresponded to optically thin synchrotron emission, 
with the mean slope $\alr \simeq 0.52$ in the range of 
$4.89  \leq \nu \leq 22.49 \,\rm GHz$. On that day the flux at 
$\nu \approx 4.89\,\rm GHz$ 
was  $S_\nu \approx 500 \,\rm mJy$ . Meanwhile, 
on November 29.97 the fluxes were at the level of $\simeq 100 \,\rm mJy$, and  
the spectrum was inverted in the range up to at least
$ 15 \,\rm GHz$. By December 6.6 the fluxes have increased further, 
reaching $S_\nu = 1083\,\rm mJy$ at 3.28 GHz. 
Comparison with the 
flux measured simultaneously by IRAM telescope at $\nu=234 \,\rm GHz$
(Rodriguez et al. 1995) shows that in this broad range the spectrum was 
`optically thin', with $\alr = 0.51$, indicating that the source during
$\leq 14$ hours became transparent at 9 cm band.

An example of gradual steepening of  initially hard spectra,  
with $\alr \sim 0.5$ at the stage of flare maximum, is provided by 
the outburst on 11 December  1993, when the fluxes  measured 
by Nancay telescope at 21 and 9 cm  have increased to 1319 and 901 mJy, 
respectively ($\alr \simeq 0.45$). 
On the next day the fluxes
have decreased by factor of 2, while the spectrum steepened to 
$\alr \simeq 0.6\,$. Gradual steepening of the radio spectra in time 
is obvious also in the multiwavelength VLA data taken 
on December 14, 16 and 17, when the spectral 
slope between 4.89 and 22.49 GHz was changing as  
$\alr \simeq 0.72, \, 0.83$ and 1.1, respectively (see Rodriguez et al. 1995). 

Similar spectral evolution, although on significantly larger
timescales, was observed for the prominent
strong radio outburst of 19 March 1994, when for the first time a pare of 
counter ejecta were resolved, and superluminal
motion in the Galaxy was discovered (MR94). On March 24  
the spectral index at $\lambda \sim 3.5\,\rm cm$ was $\alr \simeq 0.5\,$, which 
by April 16 steepened  to $\alr \simeq 0.8\,$. For an 
estimated distance $d=12.5 \pm 1.5 \,\rm kpc$ to 
GRS 1915+105 , the apparent 
velocities of the approaching and receding ejecta 
made $1.25\pm 0.15$  and $0.65\pm 0.08$ speed of light, respectively,
using which the real speed $\beta \simeq 0.92 $ of the ejecta
moving in the opposite directions at an angle $\theta \simeq 70^{\circ}$ 
to the line of sight was inferred (MR94). 

Importantly, since {\it both} 
radio emitting components of this strong and long-lived flare were  
{\it resolved}, the data of radio monitoring of this flare presented in MR94  
contain a unique information about the processes 
in the jets. Thus, the detection of the receding
component  enables measurement of the flux ratio 
$S_{\rm a}/S_{\rm r}$ of the approaching and receding ejecta. 
In the case of {\it identical} jets, the ratio of the flux densities 
measured at equal angular separations of the 
jets from the core is 
\begin{equation}
\frac{S_{\rm a}}{S_{\rm r}} = \left( \frac{1+\beta \cos\theta}
{1-\beta \cos \theta} \right)^{j+\alr}\; ,
\end{equation}
where $j=3$, if the fluxes are
produced in the moving {\it discrete} radio clouds, but $j=2$
for the brightness ratio of continuous {\it stationary } jets
(e.g. see Lind \& Blandford 1985).
 The radio patterns of the March-April flare clearly show motion of two
discrete components, but not a stationary jet emission. Then, for the deduced
values of $\beta\simeq 0.92$ and $\theta=70^\circ$, equation (1) 
predicts the flux ratio $S_{\rm a}/S_{\rm r} \simeq 12$,   
while $S_{\rm a}/S_{\rm r} = 8\pm 1$ is observed (MR94). 
This discrepancy between the observed and expected flux ratios can be 
interpreted  as an indication of shocks propagating 
with the speed $\beta \simeq 0.92 $ in the continuous fluid of 
radio emitting plasma which is steaming with a  velocity 
$\beta_{\rm f}\simeq 0.7$ 
(Bodo \& Ghisellini  1995).
However, it is possible also to explain 
(Atoyan \& Aharonian 1997) the observed 
flux ratio in terms of motion of discrete radio sources with the speeds 
coinciding with the deduced pattern speed, if one would
allow for the twin ejecta be {\it similar}, but not necessarily identical,
as implied
in equation (1). In Section 4 we will show that the `discrepancy' between 
the observed and expected flux ratios are easily removed by a small asymmetry
between the intrinsic parameters of the ejecta, e.g. by small difference in 
the velocities $\beta_{\rm a}$ and    $\beta_{\rm r}$ of the approaching
and receding clouds. Then,  deviations 
of $S_{\rm a}/S_{\rm r}$ from the `expected' value can be rather considered  
as a measure of asymmetry between the twin ejecta.

\subsection{Radio electrons}

Generally, gradual transition from self-absorbed to optically thin spectra 
at progressively lower frequencies is a typical indicator of the synchrotron 
radiation of an expanding radio cloud.
Power-law radiation spectra with $\alr \simeq 0.5$ imply power-law distribution
of relativistic electrons $N(\gamma) \propto \gamma^{-\ale}$ with 
the exponent $\ale = 2\,\alr + 1 \simeq 2$. Since these radio 
spectra span a broad interval of frequencies 
from $\sim 1\,\rm GHz$ up to $\leq 250\,\rm GHz$, the electron distribution 
should be power-law within some interval of 
energies  $\gamma_{1} \leq \gamma \leq \gamma_{2}\,$, 
with the ratio of Lorentz factors
$\gamma_{2}/\gamma_{1} >10$. To estimate the range of basic parameters 
of radio clouds, such as the magnetic field $B$, size $R$, speed of 
expansion $v_{\rm exp}$,  
etc., in this section we will suppose pure power-law electron distribution,
$N(\gamma) = A_0 \gamma^{-2}$,  extending 
from $\gamma_{1} =1$ up to $\gamma_{2} = \gamma_{\rm max}$, with a sharp 
cut-off above $\gamma_{\rm max}$, i.e. $N(\gamma > \gamma_{\rm max})= 0$. 
For this spectrum of electrons $A_0$ 
actually corresponds to the total number of relativistic electrons, 
$N_{\rm e}= A_0$, and the  
total energy in the electrons is $W_{\rm e} = A_0\, m_{\rm e}c^2 
\ln\gamma_{\rm max}$. 

Characteristic frequency of synchrotron
photons emitted by an electron with Lorentz factor $\gamma$ in the magnetic
field B is $\nu_{\rm s}\simeq 0.3\times (1.5 \nu_{\rm \,B}\, \gamma^2)$, where
$\nu_{\rm\, B}=e B \sin \vartheta /2\pi m_{\rm e}c$, and $\vartheta$ is the 
pitch angle (e.g. Ginzburg 1979). For the mean 
$\overline{\sin \vartheta} = \sqrt{2/3}$, the frequency
\begin{equation}
\nu_{\rm s} \simeq 10^6 \, B \gamma^2 \; \rm Hz\, ,
\end{equation}  
where $B$ is in Gauss.
Since the magnetic fields expected in the cloud 
$B\sim (0.03-0.3) \,\rm G$ (see
below), for production of synchrotron photons at $\nu \sim 300 \,\rm GHz$,
electrons with $\gamma \sim$ few times $10^3$ are required. In
the case of  synchrotron origin of the IR  jet 
observed by Sams et al. (1996), or of the rapid IR flares 
detected by Fender et al. (1997) in GRS 1915+105, relativistic electrons with 
$\gamma \gg 10^4$ are to be supposed. However, in the first part of this 
paper, where we study the radio fluxes,  we will not speculate with 
such possibility and will limit the range of relativistic electrons under 
consideration with $\gamma_{\rm max}=10^4$. 

\subsection{Magnetic fields, size, and speed of expansion of radio clouds}

In the case of an isotropic source with luminosity $L^{\prime}(\nu^\prime)$
in its rest frame, an observer at a 
distance $d$ would detect the flux 
$S_{\nu} =\delta^3 L^{\prime}(\nu /\delta) / 4\pi d^2$, where 
$\delta=\sqrt{1-\beta^2}/(1-\beta \cos\theta)$ is the 
Doppler factor of the source moving with a speed $\beta$ at an angle 
$\theta$ to the line of sight (e.g. Lind \& Blandford 1985). For distances 
$d_{\ast}\equiv d/12.5{\,\rm kpc} \simeq 1$,  the fluxes 
expected from an optically thin cloud are  
\begin{equation}
S_{\nu} \simeq 2 \times 10^{-39}\, A_0 \, \delta^{7/2} \, 
B^{3/2} \, \nu^{-1/2} \,d_{\ast}^{-2} \;
\rm mJy \; .
\end{equation}
Introducing the ratio $\eta= w_{\rm e}/ w_{\rm B}$ of the energy densities of 
relativistic electrons 
$w_{\rm e} = 3 W_{\rm e}/4\pi R^3$ to the energy density of the magnetic field 
$w_{\rm B}=B^2/8\pi$ in the cloud, 
the coefficient $A_0$ can be expressed as 
$A_0 = 2\times 10^{50} \,\eta \, B^2 R_{15}^3$, where $R_{15}$ is the 
characteristic radius $R$ of the cloud in units of $10^{15}\,\rm cm$. 
Since the radio fluxes observed at 
$\nu \sim 10\,\rm GHz$ reach the level  $\sim 500 \,\rm mJy$ (for strong
flares), the magnetic field in the emission region should be
\begin{equation}
B = 7.7\times 10^{-2}\,\delta^{-1}\, \eta^{-2/7} R_{15}^{-6/7} 
d_{\ast}^{4/7} S_{\ast}^{2/7} \;\rm G \;,
\end{equation}
where $S_{\ast}\equiv S_{\rm 10 \, GHz} /500 \,\rm mJy$. 

The cloud's radius $R$ can be estimated from the requirement of optical 
transparency of the source with respect to synchrotron self absorption. 
Calculation of the absorption coefficient for the
supposed power-law spectrum of electrons is straightforward 
(e.g. Ginzburg 1979), resulting in
$\kappa_{\nu} = 4.4 \times 10^{15} \eta B^{4}\nu^{-3}\,\rm cm^{-1} $. 
Using equation (4), the size $R$ of the cloud can be expressed 
in terms of the opacity $\tau_{\nu} = R\,\kappa_\nu$ at a given frequency 
$\nu$: 
\begin{equation}
R_{15} = \, 0.46 \, \delta^{-28/17}\eta^{-1/17} 
d_{\ast}^{16/17} S_{\ast}^{8/17} 
\tau_{\nu}^{-7/17} \nu_{\rm GHz}^{-21/17} \; , 
\end{equation}
where $\nu_{\rm GHz} \equiv \nu/1\,\rm GHz$. 
Then from equation (4) follows that 
\begin{equation}
B \simeq 0.15 \, \delta^{7/17} \eta^{-4/17} d_{\ast}^{-4/17} S_{\ast}^{-2/17} 
\tau_{\nu}^{6/17}\nu_{\rm GHz}^{18/17} \; \rm G .
\end{equation}

For the ejecta in GRS 1915+105 the Doppler factor 
$\delta \simeq 0.6$ (MR94), so the optical transparency 
$\tau_{\nu} \leq 1$  at frequencies $\nu \sim (1.4 - 3)\,\rm GHz$ 
requires a magnetic field at the stage of flare maximum 
$B \sim (0.2-0.3)$ Gauss, if the energy densities of relativistic electrons 
and magnetic fields in the ejecta 
would be at the equipartition level $\eta \sim 1$.

Due to rather weak dependence of $B$ on
the opacity $\tau_{\nu}$, the equipartition magnetic field $B_{\rm eq}$ 
could be significantly
less than $ 0.2\,\rm Gauss$ only if $\tau_{\nu} \leq 0.1$. However,
to reach even  $\tau_{\nu} = 1$ in a few days after ejection, the clouds 
should {\it expand with a very 
high speeds} comparable with the speed of light. 
Indeed, from equation (5) follows that
the characteristic size of the clouds ejected on 19.8 March 1994, should 
increase to 
$R\simeq   8\times 10^{14} \tau_{1.4\,\rm GHz}^{-0.41}\,\rm cm$ by 24.3 March,
$\Delta t \simeq 4.5\,\rm days$ after the ejection event, when the flare
was detected by Nancy telescope (Rodriguez et al. 1995). Then, since the 
intrinsic time 
in the rest frame of the cloud $\Delta t^{\prime} = \delta \, \Delta t$,
a speed of expansion $v_{\rm exp} > 0.1
\times c$ is to be supposed to provide $\tau_{1.4\,\rm GHz}\leq 1$.  
Expansion speeds even higher could be needed for the flares which may be 
weaker, but become transparent during $\Delta t \leq 1 \,\rm day$.
For example, for the radio flare reported by Gerard (1996),  
apparent evolution of the fluxes, measured at 9 and 21\,cm, from 
an inverted spectrum 
to the one with $\alr \simeq 0.5$  
occurred between 9.0 and 10.0 of July
1996, when the flux at 21 cm has increased from 
$S_{\rm 1.4 \,GHz} < 30\,\rm mJy$
to $S_{\rm 1.4 \,GHz} = 140 \,\rm mJy$. Using equation (5), an
expansion speed $v_{\rm exp} \geq 0.2\, c $ (!) is 
found.
Thus, transition of the flares from optically thick to thin
spectral forms on timescales of days implies that the radio clouds are 
expanding with velocities $v_{\rm exp} \geq (0.1-0.2) \, c$. 
This estimate agrees with the clouds expansion speed 
deduced from the observations, and suggests that the lack of
red-shifted optical lines from GRS 1915+105 might be due to their large 
Doppler broadening (Mirabel et al. 1997).

\subsection{Steepening of the radio spectra: implications}

The observed gradual steepening of
the radio spectra from initial $\alr \sim 0.5$ to $\alr \sim 1$ at the 
stage of fading of the flares, requires a steepening
of the energy distribution of the parent electrons from the power-law
exponent $\ale \sim 2$  
to $\ale \sim 3$. The steepening of the electron spectral index by factor 
of 1  
is usually attributed to the synchrotron (or inverse Compton) energy losses 
of the electrons,
$P_{\rm s}(\gamma) \propto \gamma^2$, and for extragalactic jets the models
of this kind were considered in a number of papers (e.g., Marscher 1980; 
K\"onigl 1981; Bloom \& Marscher 1996; etc). 
In the case of GRS 1915+105, however,
such an interpretation of the radio data is impossible. Indeed,
the synchrotron energy loss time of an electron with energy $\gamma$ is
\begin{equation}
t_{\rm s}(\gamma) 
\simeq 7.7 \times 10^8 \, \gamma^{-1} B^{-2} \; \rm s.
\end{equation}
Using equation (2), the time  
needed for modification of the electron
spectra, in the range of energies responsible for the synchrotron radiation
at frequencies $\nu$, is found:
\begin{equation}
t_{\rm s}(\nu)\simeq 2.4\times 10^7 \,\nu_{\rm GHz}^{-1/2} 
B^{-3/2} \; \rm s. 
\end{equation}
Thus, in the magnetic fields limited by equation (6), the synchrotron 
losses of the electrons  
cannot be responsible for the steepening of 
the radio spectra, since even for frequencies $\nu \sim 10 \,\rm GHz$ 
they would require timescales  larger than
few years. 
 
There are two principle possibilities to explain the observed spectral 
evolution. The first one is 
to assume that the relativistic electrons are {\it continuously} injected
into the cloud with a spectrum $Q(\gamma,t)\propto \gamma^{-\alinj}$,
with the power-law index $\alinj =2$ during the first $(1-3)$ days,
which later on steepens 
to $\alinj \geq 3$  so as to provide the steepening of the prompt
 distribution of electrons 
$N(\gamma,t)$  to the power-law index $\ale \sim 3$.  The second option is 
to assume {\it energy-dependent 
escape} of electrons from the cloud, while the injection spectrum would not 
necessarily steepen. Note that in this case as well one needs a continuous
injection  of the electrons, since otherwise for any escape-time 
$\tau_{\rm esc}(\gamma)$ the escape of electrons would 
result in a distribution $N(\gamma,t)$ with a sharp (exponential) cut-off, 
instead of power-law modification,
above some  $\gamma^{\prime}$ defined from the equation  
$\tau_{\rm esc}(\gamma^{\prime})= t$ (see Section 3). Thus, in both cases 
a proper interpretation of the spectral steepening during the flare implies
{\it continuous injection} of relativistic electrons into the clouds.

In principle, neither of those two options can 
be excluded without specification of $Q(\gamma,t)$ and detailed 
calculation of the electron distribution $N(\gamma,t)$.
However, there are some arguments in favor of the 
escape of relativistic electrons as the main reason for the
steepening of radio spectra. Indeed, let us suppose that the electron
escape is negligible. Then, in order to modify the electron 
distribution from the power-law index 
$\ale \sim 2$ at $t_1 \simeq (1-3)\,\rm days$ to 
$\ale\rightarrow 3$ at $t\sim 2\, t_1$ (as observed), the injection
of electrons with the steep spectrum $\alinj \geq 3$ 
should proceed {\it at least} with the same rate $Q(\gamma,t)$ as initially.
However, contrary to this requirement, strong steepening of the injection 
spectrum generally indicates 
that the efficiency of 
particle acceleration is essentially {\it suppressed}, so significant 
decline of the injection rate $Q(\gamma,t)$ would have to be expected. 
Meanwhile, assuming
fast escape of relativistic electrons from the cloud, the old 
`hard spectrum' electrons could be replaced with a freshly injected 
`soft' ones on timescales $\sim \tau_{\rm esc}$. Moreover,
since $N(\gamma,t) \sim Q(\gamma,t) \,  \tau_{\rm esc}$ 
(for $t\gg \tau_{\rm esc}$), 
then an {\it energy-dependent} escape, $\tau_{\rm esc}(\gamma)\propto
\gamma^{\Delta}$ with $\Delta \leq 1$, could provide the
required steepening of $N(\gamma,t)$ even for the same hard 
injection spectrum $Q \propto \gamma^{-2}$ as initially. This is an 
important possibility, since as we shall see later, the decline 
of $S_{\nu}(t)$ observed during March-April 1994 (MR94)
implies  substantial injection of relativistic electrons into the 
radio clouds at all stages of the flare evolution.

\subsection{Energetics}

The energy $W_{\rm B}$ of the magnetic 
field in the radio cloud can be estimated using equation (4):
\begin{equation}
W_{\rm B} \simeq  9.9 \times 10^{41} \,\delta^{-2} \eta^{-4/7} R_{15}^{9/7}
d_{\ast}^{8/7} S_{\ast}^{4/7} \;{\rm erg}\, . 
\end{equation} 
The energy of relativistic electrons  $W_{\rm e} = \eta W_{\rm B}$, therefore 
the total energy $W_{\rm tot} =  W_{\rm e} + W_{\rm B} \propto (\eta^{3/7}+
\eta^{-4/7})$ reaches the minimum at $\eta =4/3$, i.e. around equipartition
between magnetic field and relativistic electrons in the cloud 
(e.g. Pacholczyk 1970). For the ejecta in GRS 1915+105 with
$\delta \simeq 0.6$,  the minimum energy in radio electrons 
at the stage of flare maximum can be estimated as 
$W_{\rm e}\sim 3\times 10^{42}\,\rm erg$, which requires
continuous injection of the electrons
with the mean power $P_{\rm inj} \simeq (1-3)\times 10^{37}\,\rm erg/s$
during $\Delta t^{\prime} \sim 1-3 \,\rm days$. This 
estimate is in agreement with the one obtained by Liang \& Li (1995), 
however, it does not take into account the adiabatic energy losses of 
electrons in the rapidly expanding cloud. 
As it is shown by accurate numerical calculations in Section 4, the
adiabatic losses result in an 
increase of the electron injection needed      
to the level of $P_{\rm inj} \sim 10^{38}\,\rm erg/s$.
Note that  the power needed for bulk acceleration of the ejecta in 
GRS 1915+105 is estimated as 
$P_{\rm jet} \sim (1-5)\times 10^{38}\,\rm erg/s$ (Meier 1996).

The value of $P_{\rm inj}\sim 10^{38}\,\rm erg/s$ corresponds to the 
minimum  power of relativistic electrons to be 
injected into the cloud. In the case
of deviation of the magnetic field $B$ in the cloud from its `equipartition' 
value $B_{\rm eq}$,  
the energy requirements would be significantly higher. 
Defining $B_{\rm eq}$ as a magnetic field for which $\eta = 1$, 
from equation (6)  follows that $\eta = (B/B_{\rm eq})^{-17/4}$. 
It means that a decrease of $B$ by an order 
of magnitude is to be compensated by an increase of $\eta$ 
by {\it four} orders of magnitude, and correspondingly 
the increase of $W_{\rm e}\propto \eta^{3/7}$ by almost 2 orders of 
magnitude, to provide the same flux $S_\ast$.
Thus, assuming a magnetic field $B_0\sim 0.03 \,\rm G$  at times 
$t \sim 3 \,\rm days$ after
ejection  (when the cloud's radius $R\sim 10^{15} \,\rm cm$), instead of
$B_{\rm eq} \simeq (0.2-0.3)\,\rm G$,  
one has to suppose  
$P_{\rm inj} \sim 10^{40} \,\rm erg/s$, which significantly 
exceeds the maximum luminosity of GRS 1925+105 observed in the X-rays
(Greiner et al. 1996; Harmon et al. 1997).
So, $B_0\geq 0.03$ should be considered as 
a rather conservative lower limit for the characteristic 
magnetic field in the radio clouds at  stages of flare maximum.

Increasing magnetic fields by an order of magnitude 
beyond equipartition field $B_{\rm eq}$, we would decrease $P_{\rm inj}$
down to $10^{36} \,\rm erg/s$. But now the energy in the magnetic field
would strongly increase, so in order to account for these 
magnetic fields, we would have to suppose 
a total power in the jet $P_{\rm jet} \geq 10^{40}\,\rm erg/s$. 
Moreover, for the fields exceeding $B_{\rm eq}$ even by factor
of 2, the magnetic pressure in the cloud would significantly {\it exceed} 
the one of the relativistic electrons,
in which case one would have to explain why a strongly magnetized clouds are
expanding with sub-relativistic speeds (in the opposite case of $\eta
\ll 1$ the answer is obvious).

For the estimated energy $W_{\rm e}\simeq 3\times 10^{42} 
\eta^{3/7} \,\rm erg$, and the supposed spectrum $N(\gamma) = A_0  
\gamma^{-2}$ in the range $1\leq \gamma \leq 10^4$, 
the total number of relativistic electrons
in the cloud is $N_{\rm e}= A_0 \simeq 4\times 10^{47}\eta^{3/7} $. 
Assumption of equal amount of  protons results in the cloud's mass 
$M_{\rm cl} \simeq 7\times 10^{23}\eta^{3/7} \,\rm g$. 
In the equipartition
state this estimate is by an order of magnitude smaller than the one 
in Mirabel \& Rodriguez (1995). Note, however, that the magnetic field 
used by Mirabel \& Rodriguez is $B\simeq 0.05 \,\rm G$,
which is by factor $\sim 5$ smaller than  
$B_{\rm eq}$. For these magnetic fields, which cannot be 
excluded (moreover, seem quite possible, see below), the parameter 
$\eta \sim 10^3$, resulting in estimated mass 
$M_{\rm cl} \simeq 10^{25}\,\rm g$.

The kinetic energy of a cloud moving with the bulk 
Lorentz-factor $\Gamma \simeq 2.5$ is estimated as   
$W_{\rm kin}= (\Gamma-1) M_{\rm cl} c^2 \simeq 10^{45}\eta^{3/7} \,\rm erg$,
i.e. essentially more than $W_{\rm e}$. This is explained by the fact
that the mean energy per relativistic electron is $\overline{\gamma}=
\ln{\gamma_{\rm max}}\simeq 10$, while the assumtion of equal amount of
 protons
in the cloud immediately increases the energy by factor of 
$(\Gamma-1) m_{\rm p}/m_{\rm e} \overline{\gamma} \simeq 300$. 
Then for acceleration of the ejecta in a day or 
shorter timescales, a power $P_{\rm jet}
\gg 10^{40}\,\rm $ would be needed. Obviously, this rather uncomfortable 
requirement significantly softens if one assumes $e^{+}-e^{-}$ pair plasma  
(e.g. Liang \& Li 1995; Meier 1996). Note, however, that it is quite
possible to reduce the required power in the jet down to the level 
$\sim 10^{38}\,\rm erg/s$, while still assuming electron-proton plasma 
in the ejecta. Indeed, if the spectrum of the electrons flattens below
some $\gamma_1 \leq 100$ (for $B\sim 0.1 \,\rm G$ the synchrotron radiation 
of these electrons is below GHz domain), then the total 
number of electrons and protons 
may be  2 orders of magnitude less than 
estimated above, resulting in the proportional reduction of the  
estimated kinetic energy $W_{\rm kin}$ of the ejecta.

\subsection{Evolution of the magnetic field in time}

An important information about the evolution of basic parameters  
in the radio clouds is found from the analysis 
of the observed rate of decline
of the flare $S_{\nu}(t) \propto t^{-1.3}$ during 
March-April 1994 (MR94).
Generally, $S_{\nu}(t)$ depends on the temporal evolution of the 
total number of radio electrons and of the magnetic field $B(t)$. 
Conclusions concerning possible behavior of $B(t)$ 
could be inferred from the condition 
of 'freezing' of the magnetic 
field lines into the highly conductive fluid. This implies that 
$ B/\rho \propto \Delta l$, where $\rho$ is the plasma  
mass density, and ${\Delta l}$ is a fluid element along the magnetic field 
line (see Landau \& Lifshitz 1963). The plasma density depends on time as
$\rho(t)\propto M_{\rm cl}(t)/[R(t)]^{3}$. 
Assuming now that: 

\noindent
(a) there is no strong 
turbulent eddies in the cloud, so the length of a fluid element in the 
expanding medium scales as $\Delta l(t) \propto R(t)$; {\it and}

\noindent
(b) the mass of the cloud is constant, 

\noindent
the magnetic field would be $B\propto
R^{-2}\propto t^{-2}$, as long as $R\simeq v_{\rm exp}\, t$. 
Then from equation (3) follows that the radio flux will drop as $S_{\nu} 
\propto N_{\rm e}(t) t^{-3}$. Therefore, to account for the observed 
decline of radio flux, one would have to assume an increase of the total number
of relativistic electrons as fast as $N_{\rm e}\propto t^{1.7}$. Even
neglecting adiabatic energy losses,  such a behavior of $N_{\rm e}$ 
implies in-situ acceleration/injection of the electrons with the rate {\it 
increasing} as $Q(t)\propto t^{0.7}$, resulting in very hard requirements 
to be imposed on the jet energetics at the late stages. Moreover, the
dependence of $B\propto R^{-2}$ results in $W_{\rm B}\propto R^{-1}$,
so the total energy of the magnetic field would be extremely high,
$W_{\rm B} > (10^{44}-10^{46}) \,\rm erg$,
if extrapolating backward from $W_{\rm B} \sim 10^{42} \,\rm erg$ at the 
stage of flare maximum to earlier stages\footnote{We remind that
$R_{15}$ in Eqs.~(4) and (5) relates to the size of an {\it optically thin} 
cloud from which unabsorbed flux $S_{\ast}$ is detected, 
therefore these equations cannot be used
for estimation of $B$ at initial stages.}, 
when $R\ll 10^{15}\,\rm cm$.  
 
To avoid these problems, we have to suppose that at least one of the 
assumptions (a) or (b) above does not hold. If a turbulence is developed
in the plasma, then in addition to stretching due to cloud's radial expansion,
the fluid lines will be stretched in turbulent eddies, so $\Delta l 
\propto R^{1+a}$ with $a > 0$, resulting in $B(R)\propto
R^{-2+a}$ (a slower decline due to turbulent dynamo effect).
If the mass of the cloud is not constant, but rather is accumulated in
time as $M_{\rm cl} \propto t^{b}$,  
then $B\propto t^{-2+b}$ (additional `compression' effect).
So, the magnetic field can be approximated as $B\propto R^{-m}
\propto t^{-m}$,
and an index $m < 2$ would indicate that the magnetic fields
are effectively created/supplied in/into the radio clouds. Since $W_{\rm B}
\propto R^{3-2m}$, a power-law index $m < 1.5$ will eliminate the energetical 
problems for initial stages of cloud's evolution. On the other hand, 
in the case of $m < 1$  similar problems  appear again, 
but now for later stages when $R\gg 10^{15}\,\rm cm$.
Thus, the range of $1 \leq m < 1.5$ seems to be most reasonable for this
parameter. Then the observed decline $S_{\nu}(t)\propto t^{-1.3}$  
can be qualitatively explained, assuming that 
the total number of radio electrons in the cloud increases as
$N_{\rm e}\propto t^{n}$ with an index $n \simeq 1.5 m -1.3 > 0$. 
Even neglecting adiabatic and escape losses of the electrons,
one would need continuous injection of radio electrons with    
$Q(\gamma,t)\propto t^{\,n-1}$, which means that the injection 
may be stationary or gradually decreasing, 
but {\it substantial} at all stages of the flare.

\subsection{The principal scenario}

Summarizing this Section, the following scenario for the  
radio flares in GRS 1915+105 can be proposed.  At the first stages of the
outburst, the radio emitting plasma forming a pair of radio clouds is 
ejected at relativistic speeds from the  vicinity of the
compact object, presumably, a black hole in the binary. 
While moving at  relativistic
speeds $\beta \simeq 0.9$ (at least, for the 19 March 1994 event)
in the opposite directions, the  
twin clouds (with similar, but not necessarily identical parameters)
are also {\it expanding} with a high speeds, $v_{\rm exp} 
\sim (0.1-0.2) c$, to reach a size $R \sim (0.5-1) \times 10^{15} \,\rm cm$ 
and become optically transparent for synchrotron self-absorption  in
a few days after expulsion. At that times the equipartition magnetic field
in the clouds $B_{\rm eq} \sim (0.2-0.3)\,\rm G$ is expected, 
but it may be also several times smaller, 
which implies a strong impact of the electrons on the dynamics of 
the expanding clouds.   
Relativistic electrons are continuously injected into the emission region
with a spectrum $Q(\gamma,t) \sim \gamma^{-2}$, and characteristic
power $P_{\rm inj}\sim 10^{38} \,\rm erg/s$ or more, depending on
the magnetic field $B\leq B_{\rm eq}$.
These electrons may be due to, e.g., in-situ acceleration at the bow shock
front ahead of the cloud, or a relativistic wind of magnetized plasmas
 propagating in the jet region,  and pushing the clouds
forward against the ram pressure of the external medium. An important
point is that, simultaneously with injection, the electrons  also escape
from the cloud on energy-dependent timescales $\tau_{\rm esc}(\gamma)$, which
modifies the electron distribution $N(\gamma,t)$, and explains the 
steepening of the radio spectra of the fading flare to $\alr \sim 1$. 
The decline of the flare  
is due to combination of decreasing magnetic field in the expanding cloud,
decline of the injection rate $Q(\gamma,t)$, as well as adiabatic 
energy losses of radio electrons and their escape from the cloud.

\section{Relativistic electrons in an expanding medium}

\subsection{Kinetic equation}

For calculations of the synchrotron and inverse Compton fluxes 
expected in the framework of above described
scenario, we have to find energy distribution function $N\equiv N(\gamma,t)$ 
of the electrons in this essentially nonstationary conditions. The 
kinetic equation describing the evolution of relativistic electrons which we 
consider in this section, is a well known partial differential equation  
(e.g. Ginzburg \& Sirovatskii 1964):
\begin{equation}
\frac{\partial N}{\partial t}\, = \, \frac{\partial}{\partial \gamma}(P N)
\, -\, \frac{N}{\tau} \, +\, Q\; .
\end{equation}
The Green's function solution to this equation in the case of 
time-independent energy losses and constant escape time 
$\tau(\gamma,t)= const$  was 
found by Syrovatskii (1959). However, in an expanding magnetized cloud 
under consideration we have to suppose that {\it all} parameters 
depend on both energy $\gamma$ and time $t$, i.e. $Q\equiv
Q(\gamma,t)$ is the injection spectrum, $\tau \equiv \tau(\gamma,t)$ is
characteristic escape time of a particle from the source, and $P\equiv
P(\gamma,t)= -(\partial \gamma/\partial t)$ is the energy loss rate.

Strictly speaking, equation (10)  
corresponds to a spatially homogeneous
source where the energy gain due to 
in-situ acceleration of particles is absent. Actually, however, it
has much wider applications, and in particular, it seems to be quite sufficient
in our case. Indeed, in a general form the equation describing 
evolution of the local (i.e. at the point {\bf r}) 
energy distribution function $f\equiv f(\gamma,{\bf r}, t)$
of relativistic particles  can be written as
(e.g. Ginzburg \& Syrovatskii 1964):
\begin{equation}
\frac{\partial f}{\partial t} = {\rm div} ({\sf D}_{\rm r}
{\bf grad} f) -
{\rm div}({\bf u}_{\rm r} f)+ \frac{\partial}{\partial \gamma}(P_{\rm r} f)
- \frac{\partial}{\partial \gamma}(b_{\rm r} f)
 +\frac{\partial^2}{\partial \gamma^2}(d_{\rm r} f) \; ,
\end{equation}
where all parameters depend also on the radius-vector {\bf r}. First two 
terms in the right side of this equation describe diffusive and
convective propagation of particles, the last two terms correspond to 
the acceleration of the particles through the first and second order 
Fermi mechanisms. If there are internal sources of particle 
injection and sink (such as production and annihilation), then the 
terms similar to
the last two ones in equation (10) should be added as well.

Let us consider a source where the region of effective 
particle acceleration can be separated from the main 
emission region. It seems to be the case for radio clouds in GRS 1915+105,
where probable site for in-situ acceleration of electrons  is a 
relatively thin region  around either the bow shock front formed ahead of 
the cloud,
or possibly a {\it wind termination} shock formed behind of the cloud,  
in the contact of the 
relativistic wind in the jet region with the cloud, as discussed below
in this paper.
Meanwhile the main part of the observed flux should be produced in 
a much larger volume $V_0$ of the post-shock region in the cloud, 
since the synchrotron cooling time of radio electrons  
is orders of magnitude larger than the dynamical times of the source. 
Since acceleration efficiency (parameters $b_{\rm r}$ and $d_{\rm r}$) 
should significantly drop outside of the shock region, after 
integration of equation (11) over the volume $V_0$ the last two terms can be 
neglected. 

The integration of the left side of equation (11) results 
exactly in 
$\partial N/{\partial t}$. Integration of the two propagation terms in 
the right side of equation (11) gives the net flux of particles, due to
diffusion and convection, across the surface of the emission region. 
Obviously, these terms are expressed as the difference between the total 
numbers of particles 
injected into and escaping from the volume $V_0$ per unit time, 
so the last two terms of equation (10)
are found (internal sources and sinks, if present, are also implied).
At last, integration of the energy loss term in equation (11) is reduced to the 
relevant term of equation (10), 
where $P$ corresponds to the mean energy loss rate
per a particle of energy $\gamma$, 
i.e. $P=\int{P_{\rm r} f {\rm d}{^3}r}/\int{f{\rm d}{^3}r}$. Under the volume
$V_0$ of the cloud 
we understand the region, filled with relativistic electrons and {\it enhanced}
magnetic field, where the bulk of the observed 
radiation is produced. 
 
Thus, equation (10) is quite applicable for the study of 
sources with ongoing in-situ acceleration, 
as long as the volume $V_0$, where the bulk of nonthermal radiation is 
produced, is much larger than the volume 
$\Delta V$ of the region(s) in the source responsible for effective 
acceleration of the electrons. 
We should mention here that solutions for a large number of particular cases
of the Fokker-Planck partial differential equation (which includes 
the term $\propto \bar{d_{\rm r}}$ for stochastic acceleration), 
corresponding to different combinations of terms responsible for time-dependent 
adiabatic and synchrotron energy losses, stochastic and regular acceleration,
have been long ago obtained and 
qualitatively discussed by Kardashev (1962). However, only solutions 
for {\it energy-independent} escape of relativistic particles were considered,
while in our study it is a key feature for proper description of the  
spectral evolution of the radio flares. Another important 
point is that  the Fokker-Planck equation
generally may contain singularity,
so transition from  the solutions of that equation (if known), which 
are mostly expressed through special functions, to the case of 
$\bar{d_{\rm r}} \rightarrow 0$ may not always be straightforward 
(for comprehensive discussion of the problems 
related with singularities in the Fokker-Planck 
equation, as well as general solutions for {\it time-independent} parameters
see Park \& Petrosian 1995, and references therein).
Meanwhile,  substitution of the acceleration terms by effective
injection terms in the regions responsible
for the bulk of nonthermal radiation, allows us to disentangle the 
problems of acceleration and emission of the electrons, 
and enables   
analytical solutions to the first order equation (10) which are 
convenient both for further qualitative analysis and 
numerical calculations.

\subsection{ Time-independent energy losses}

Suppose first that the escape time is given as $\tau=\tau(\gamma,t)$, but
the energy losses are time-independent, $P=P(\gamma)$. The Green's function
solution $G(\gamma, t, t_0)$ to equation (10)  
for an arbitrary injection spectrum 
$ N_{0}(\gamma)$ of
electrons implies $\delta$-functional injection 
$Q(\gamma,t) = N_{0}(\gamma)\,\delta(t-t_0)$ at some instant $t_0$. At 
times $t>t_0$ it actually corresponds to the solution for the homogeneous
part of equation (10), with initial condition 
$G(\gamma, t_0+0, t_0) =  N_{0}(\gamma)$, while 
$G(\gamma, t_0-0, t_0)=0$.    
Then for the function $\,F\,=\,P G\,$ this equation is  
reduced to the form
\begin{equation}  
\frac{\partial F}{\partial t} = \frac{\partial F}{\partial \zeta} -
\frac{F}{\tau_{1}(\zeta,t)} \; ,
\end{equation}
if introducing, instead of the energy $\gamma$, a new variable
\begin{equation}
\zeta = g(\gamma) \equiv \int_{\gamma_{\ast}}^{\gamma}{\frac{{\rm d}\gamma_1}
{P(\gamma_1)}}\; 
\end{equation}
where $\gamma_{\ast}$ is some fixed energy. Formally, $\zeta$ has a
meaning of a time needed for a particle with energy $\gamma$ to cool down
to energy $\gamma_{\ast}$ (so for convenience one may 
suppose $\gamma_{\ast}=1$).
The function $\tau_{1}(\zeta,t)= \tau[\varepsilon(\zeta),t]$, 
where $\varepsilon$
is the inverse function to $ g(\gamma)$ which
 expresses the energy through $\zeta$, i.e. 
$\gamma= \varepsilon(\zeta)$. 
The initial condition for $F(\zeta,t)$ reads 
\begin{equation}
F(\zeta,t_0) = P[\varepsilon(\zeta)]\, N_{0}[\varepsilon(\zeta)]
\equiv U(\zeta) \; .
\end{equation}  
Transformation of equation (12) from variables 
$(\zeta, \,t)$ to $(s=\zeta +t,\,u=t)$ results  
in a partial differential equation over only one variable for 
the function $F_1(s,u)=F(\zeta,t)$:
\begin{equation}
\frac{\partial F_1}{\partial u} = - \frac{F_1}{\tau_{1}(s-u,u)} \; , 
\end{equation}
with the initial condition $F_{1}(s,u_0) = U(s-u_0)$ found from equation (14).
Integration of equation (15) is straightforward:
\begin{equation}
 F_{1}(s,u)= U(s-u_0)  
 \, \exp\left[ - \int_{u_0}^{u} \frac{{\rm d} u_1}
 {\tau_{1}(s-u_1,u_1)}\right]\; 
 \cdot
 \end{equation}
In order to come back from variables $(s,\, u)$ to $(\gamma, \,t)$,
it is useful to understand the meaning of the function 
$\varepsilon(s-x)$ which 
enters into equation (16) via equation (14) for $U$  and the escape 
function $\tau_1\rightarrow \tau$. 
Since $\varepsilon $ is the 
inverse function to $g$, then for any $z$ in the range of definition
of this function we have $z=g[\varepsilon(z)]$. Then, 
taking into account that $s=\zeta +t$ and $\zeta=g(\gamma)$, for $z=s-x$ 
we obtain $t-x = g[\varepsilon(s-x)] - g(\gamma)$. For the function $g$ 
defined by equation (13), this equation results in
\begin{equation}
t-x=\int_{\gamma}^{\Gamma_{x}(\gamma,t)}{\frac{{\rm d}\gamma_1}
{P(\gamma_1)}} \;
\end{equation}
where $\Gamma_{x}(\gamma,t)$ corresponds to $\varepsilon(s-x)$   
after its transformation to the variables $(\gamma,t)$. 
Thus, for a particle with energy $\gamma$ at an instant $t$, 
the function $\varepsilon(s-x)$ is the energy $\Gamma_x \equiv 
\Gamma_{x}(\gamma,t)$ of that particle 
at time $x$, i.e. it describes the trajectory of individual particles
in the energy space.

Expressing equation (16) in 
terms of the Green's function  $G=F/P$, the solution 
to equation (10) for an arbitrary $\tau(\gamma,t)$, but 
time-independent energy losses of particles is found:
\begin{equation}
G(\gamma,t,t_0) = \frac{P(\Gamma_{t_0}) N_{0}(\Gamma_{t_0})}
{P(\gamma)} \exp \left[- \int_{t_0}^{t} \frac{{\rm d} x}{\tau(\Gamma_{x},x)}
\right] \cdot
\end{equation}
Note that this is not a standard Green's function in the sense that 
the injection spectrum was supposed as an arbitrary function of energy 
$N_{0}(\gamma)$,
and not necessarily a delta-function. Actually, it describes the 
evolution of relativistic particles with a given distribution
$N_{0}(\gamma)$ at $t=t_0$. The solution for an arbitrary continuous 
injection spectrum is readily found after substitution 
$N_{0}(\gamma) \rightarrow
Q(\gamma,t_0) {\rm d} t_0$ into equation (18) and 
integration over ${\rm d} t_0$:
\begin{equation}
N(\gamma,t)= \frac{1}{P(\gamma)} \int_{-\infty}^{t}{P(\Gamma_{t_0}) 
Q(\Gamma_{t_0},t_0)
\exp \left[- \int_{t_0}^{t} \frac{{\rm d} x}{\tau(\Gamma_{x},x)}
\right] }\, {\rm d}t_0 \; ,
\end{equation}
with the function $\Gamma$ defined via equation (17). In the 
particular case of
time and energy independent escape, $\tau(\gamma, t) = const$, this solution
coincides with the one given in Syrovatskii (1959) and Ginzburg \&
Syrovatskii (1964) in the form of double integral over $t_0$ and $\Gamma$,
if equation (19) is integrated over energy
with the use of general relations 
\begin{equation}
\frac{\partial \Gamma_{x}}{\partial t} =
- \frac{\partial \Gamma_{x}}{\partial x}= P(\Gamma_x) \;,
\end{equation} 
which follow from equation (17). 

It worths brief discussion of some specific cases of equation (19).
Let the escape of particles be energy dependent but stationary, 
$\tau(\gamma,t) \rightarrow \tau(\gamma)$, and consider first 
the evolution of $N(\gamma,t)$ when the energy losses are negligible,
so $\Gamma_x \simeq \gamma$ for any $t$. 
Assuming for convenience that the form of injection
spectrum does not change in time, i.e. $Q(\gamma,t)=Q_{0}(\gamma) q(t)$
with $q(t<0)=0$ (i.e. injection starts at t=0), equation (19) is reduced to
\begin{equation}
N(\gamma,t)= Q_{0}(\gamma) \tau(\gamma) \int_{0}^{t/\tau(\gamma)}
{q[t-\tau(\gamma) z]
\, e^{-z} \,{\rm d} z}\;.
\end{equation}
For stationary injection $q(t\geq 0) =1$ the integral results in simple
$(1-e^{-t/\tau})$, so $N(\gamma,t)\simeq
Q_0(\gamma) \, t$ until $t <\tau(\gamma)$, and then the escape 
of electrons
modifies the particle distribution, compared with the injection spectrum,
as $N(\gamma,t)\simeq Q_0(\gamma) \, \tau(\gamma)$. In the case of 
$\tau(\gamma) \propto \gamma^{-\Delta}$ it results in a power-law
steepening of the injection spectrum by factor of $\Delta$. 
This is a well known result of the leaky-box models in the theory of 
cosmic rays. In the case of nonstationary injection, however, the 
modification of $Q(\gamma)$ is 
different. In particular, for an impulsive injection $q(t)=\delta(t)$
it is reduced to a sharp cut-off of an 
exponential type above energies $\gamma_t$ found from 
$\tau(\gamma_t)\simeq t$. 
We remind, that this qualitative feature has been used in the previous 
Section to argue that for interpretation of the steepening of radio spectra
during the flares by escape, one needs continuous injection of electrons. 

For a  stationary injection of particles,  
 equation (19) can be transformed to the form
\begin{equation}
N(\gamma,t)=\frac{1}{P(\gamma)} \int_{\gamma}^{\Gamma_0} {Q_0(\Gamma) 
\exp \left[- \int_{\gamma}^{\Gamma}\frac{{\rm d} z}{P(z) \tau(z) }
\right]   {\rm d} \Gamma } 
\end{equation}
using equation (20). In the case of $\tau \rightarrow \infty$ (absence
of escape) and large $t$, when $\Gamma_0\equiv \Gamma_0(\gamma, t)\rightarrow
\infty$,  equation (22) comes to familiar 
steady state solution for distribution of particles in an infinite  medium.
If the synchrotron (or IC) 
energy losses of electrons dominate, $P=p_2 \gamma^2$, then 
$\Gamma_0 = \gamma/(1-p_2 t \gamma)$. In this case
$N(\gamma,t)\sim t\, Q_0(\gamma)  $ until $p_2 t \gamma \leq 1$, and then
 the radiative losses  result in a quick steepening of a 
{\it stationary } power-law  injection spectrum by a factor of 1.
Meanwhile, in the case of impulsive injection the modification of 
the initial 
spectrum of electrons is reduced to an exponential cut-off at 
$\gamma \geq 1/p_2 t$ (see Kardashev 1962).

\subsection{Expanding cloud}

Energy losses of relativistic electrons in an expanding medium 
become time-dependent. For the radio electrons in
 GRS 1915+105  only adiabatic losses due to expansion of the cloud
 are important. The adiabatic energy loss rate is given
as  $P_{\rm ad} = v\gamma /R $, where $R$ is the characteristic radius of the 
source, and $v$ is the speed of spherical expansion   
 (e.g. Kardashev 1962). For the electrons of higher energies, however,
the synchrotron losses  may dominate.
For the magnetic field we  suppose 
$B=B_0 (R_0/R)^{-m}$, where $B_0$ and $R_0$ are the magnetic field and the 
radius of the cloud at the instant $t_0$.  
Thus, the total energy  losses can be written as  
\begin{equation}
P\, = \frac{\gamma}{R}\left(p_1 \, +\, p_{2}\frac{\gamma}{R^{\mu}}
\right) \; ,
\end{equation}
where  $\mu= 2m-1$. For adiabatic
losses $p_1 = v$, but we keep the constants $p_1$ and $p_2$ in the parametric
form in order to enable other losses with similar dependence 
on $\gamma$ and $R$ as well.
Here we will suppose that expansion speed $v=\rm const$, and consider evolution
of  the particles injected impulsively at instant $t_0$ with the
spectrum $G(\gamma,t_0) =N_{0}(\gamma)$, as previously. 

Since the energy losses depend
on time via the radius $R(t)= R_0+v (t-t_0)$, 
it is convenient to pass from variable $t$ to 
$R$. Then for the function $\Phi = \gamma G(\gamma,R)/R$ 
equation (10) reads:
\begin{equation}
R \frac{\partial \Phi}{\partial R} = \gamma \frac{\partial}{\partial \gamma}
\left[ \left( a_1 +a_2\frac{\gamma}{R^{\mu}}\right) \Phi \right]
 \,- \, \left(1 +\frac{\gamma}{v \, \tau}\right) \Phi \; ,
 \end{equation}
where $a_1=p_1/v$, $a_2=p_2/v$, and for $\tau$ now we imply the function
$\tau(\gamma, R)$. Transformation of this equation from
variables $(\gamma,R)$ to $(\psi = \ln(\gamma/R^{\mu}), \, \xi=\ln R)$
results in the equation
\begin{equation}
 \frac{\partial \Phi_1}{\partial \xi} = 
 \frac{\partial}{\partial \psi}
 [ (\mu+ a_1 +a_2 e^{\psi}) \Phi_1 ] \, -\, 
 \left[1 +\frac{e^{\xi}}{v \, \tau_1(\psi,\xi)}\right] \Phi_1 \; ,
\end{equation}
where $\Phi_1 \equiv \Phi_1 (\psi,\xi)$ and $\tau_1(\psi,\xi) =
\tau(e^{\psi+\mu \xi}, e^{\xi})$. The initial condition at $\xi_0=
\ln R_0$ reads 
$\Phi(\psi,\xi_0)=R_{0}^{\mu-1} e^{\psi} N_0(R_{0}^{\mu}  e^{\psi})$.
Thus, we come to the equation formally coinciding with the 
one considered above, with `time' ($\xi$) independent `energy' ($\psi$) 
losses
$P_{\ast}(\psi) = \mu + a_1 +a_2 e^{\,\psi}$, and arbitrary `escape'
function 
$\tau_{\ast}(\psi,\xi) = (1+e^{\xi}/v\,\tau_1)^{-1}$. 
The solution to this equation is analogous to equation (18): 
\begin{equation}
\Phi_1(\psi,\xi) = R_{0}^{\mu-1}\,e^{\Psi_{0}}
\frac{1+c_\ast e^{\Psi_{0}}}{1+c_\ast e^{\psi}} N_{0}(R_{0}^{\mu}  
e^{\Psi_{0}}) \exp \left[\xi_{0}-\xi - \int_{\xi_0}^{\xi}
\frac{e^{z} {\rm d} z}{v\,\tau_{1}(\Psi_{z} , z)} \right] \;\; , 
\end{equation}
where $c_{\ast} = a_{2}/(\mu+a_1)$, $\Psi_0 \equiv \Psi_{\xi_0}$, 
and $\Psi_x \equiv \Psi_{x}(\psi,\xi)$ 
is  
characteristic trajectory of a particle in the `energy' space $\psi$,
which is readily calculated from equation (17) for the given $P_\ast$:
\begin{equation}
\Psi_{x}(\psi,\xi)= -\ln [(c_\ast + e^{-\xi}) e^{(\mu+a_1)(\xi - x)}-c_\ast]
\end{equation}
Returning now to the variables $\gamma$ and $R$, the evolution of  
 particles with the energy distribution $N_0(\gamma)$ at the instant $R=R_0$
is found:
\begin{equation}
G(\gamma,R,R_0)=\left(\frac{R_0}{R}\right)^{a_1}
\frac{\Gamma_{R_0}^2}{\gamma^2}
N_0(\Gamma_{R_0} )
\exp\left[- \frac{1}{v} \int_{R_0}^{R}\frac{{\rm d} r}{\tau
(\Gamma_{r}, r)} \right] \; \cdot
\end{equation}
The energy $\Gamma_{r}\equiv \Gamma_{r}(\gamma,R)$ corresponds to the 
trajectory of a particle
with given energy $\gamma$ at the instant $r=R$ in the $(\Gamma,r)$ plane, and
can be represented as $\Gamma_{r}=\gamma \, \Lambda(\gamma,R,r)$ 
where  
\begin{equation}
\Lambda(\gamma,R,r) = 
\frac{ \left(\frac{R}{r}\right)^{a_1}}
{1+\frac{c_\ast \,\gamma}{R^{\mu}}\left[1 - \left(\frac{R}{r}\right)^{\mu+a_1}
\right]} \;\cdot
\end{equation} 
In the formal case of $\mu +a_1 \rightarrow 0$, 
when $c_\ast =a_2/(\mu +a_1)\rightarrow \infty$, equation (29) tends to 
the limit $\Lambda^{\prime} =(R/r)^{a_1}/[1+a_2\gamma R^{a_1}\ln (R/r)]$.
Note that for large $\gamma$ the radiative losses may limit the 
trajectory of relativistic electrons at some $r_{\ast} \geq R_0$, when 
$\Lambda(\gamma,R,r_\ast) \rightarrow \infty$. 
For these energies $\Gamma_{R_0}$ in equation (28) should be taken as 
$\infty$, but 
$G(\gamma,R,R_0) = 0$ as far as $N_0(\infty ) = 0$.

In the general case of continuous injection of relativistic particles 
with the rate $Q(\gamma,t)\rightarrow Q(\gamma,R)$, the evolution of their 
energy distribution during expansion of the cloud between
radii $R_0$ and $R\geq R_0$ is  found, using equation (28):
\begin{equation}
N(\gamma,R)=G(\gamma,R,R_0)+\frac{1}{v} \int_{R_0}^{R}
\left(\frac{r}{R}\right)^{a_1}\Lambda^{2}(\gamma,R,r) Q(\Gamma_r,r)
\exp\left[ -\frac{1}{v}\int_{r}^{R}\frac{{\rm d} z}{\tau(\Gamma_z,z)}
\right] {\rm d} r \; .
\end{equation} 
Here $G(\gamma,R,R_0)$ given by equation (28) describes the evolution
of those particles which have already been in the cloud by the instant 
$R_0$, while the second term
describes the particles being injected during the time interval $(t_0,t)$.  
The substitution $R=R_0 +v (t-t_0)$ results in an explicit expression for 
$N(\gamma,t)$. If the source is expanding with
a constant velocity $v$ starting from $t=0$, such a substitution 
results in formal changes $R\rightarrow t$, $R_0 \rightarrow t_0$, 
$r\rightarrow t^{\prime}$, and  ${\rm d}r = v 
{\rm d} t$ in equation (30). 
If only the adiabatic losses are important, i.e $a_1 = 1$ and
$a_2 =0$, equation (29) is reduced to a simple $\Lambda = t/t_0\,$,
and 
\begin{eqnarray}
N(\gamma,t)& = &\frac{t}{t_0}\, N_0\left( \frac{t}{t_0}\gamma\right)
\exp \left(-\int_{t_0}^{t}
\frac{{\rm d} x}{\tau(t\gamma/x, x)}\right) \, + \nonumber \\
  & & \int_{t_0}^{t}\frac{t}{z}\, Q\left(
  \frac{t}{z}\gamma, z\right) \exp
\left(-\int_{z}^{t}
\frac{{\rm d} x}{\tau(t\gamma/x, x)}\right) {\rm d} z \; .
\end{eqnarray} 

For the {\it energy-independent} escape, $\tau(\gamma,t)=
\tau(t)$, a similar equation can be obtained from the 
relevant Green's function solution 
found by Kardashev (1962) in his case "stochastic acceleration + adiabatic
losses + leakage" (the single combination of terms in that work where the 
escape term was included), if one tends the acceleration parameter to zero.    

It is seen from equation (31) 
that for a power-injection $Q(\gamma)\propto \gamma^
{-\alinj}$ with $\alinj > 1$ the contribution of the first term quickly 
decreases, so at $t\gg t_0$ only contribution due to continuous injection
is important. This term is easily integrated assuming stationary injection 
and approximating $\tau = \tau_0 \gamma^{-\delta} (t/t_0)^s$, 
with $\delta$ and $s \geq 0$. In the
case of $s < 1$, the energy distribution of electrons at  
$t \gg \tau(\gamma,t)$
   comes to $N(\gamma,t)=Q(\gamma) \times \tau(\gamma,t)$, similar to the case
of non-expanding source. 
If $s\geq 1$, the condition $t \gg \tau(\gamma,t)$ 
can be satisfied only
for large enough $\gamma$, so only at these energies the energy-dependent 
escape of particles from an expanding cloud  
can result in a  steepening of $N(\gamma,t)$.  

At the end of this section we remind that equation (30) is derived under assumption
of a constant expansion speed $v$. However, 
it can be readily used in the numerical calculations for any profile of 
$v(t)$, if approximating the latter in the form of step functions with 
different mean speeds $\bar{v_{\rm i}}$ in the succession
of intervals $(t_{\rm i},t_{{\rm i} +1})$. Similar approach can be 
implemented, if needed, also for modelling of an arbitrary profile of the 
magnetic field $B(R)$.

\section{Modelling of the March 19 radio flare} 

In the framework of the scenario qualitatively described in Section 2, and
using the results of Section 3, in this Section 
we study quantitatively the  
time evolution of the fluxes of 19 March 1994 radio flare  
of  GRS 1915+105 detected by Mirabel \& Rodriguez (1994) and Rodriguez et 
al (1995). This prominent flare presents 
particular interest from several aspects: 

\noindent
({\bf a}) until now it remains a unique outburst of GRS 1915+105  
where both ejecta have been clearly resolved in VLA data and parameters
of superluminal ejecta calculated (MR94); 

\noindent
({\bf b}) it was very long-lived flare observed during $> 40\,\rm days$;  

\noindent
({\bf c}) it was very strong outburst, 
with the reported accuracy of flux measurements $\simeq 5\,\%$.  

\noindent
({\bf d}) time evolution of the fluxes from {\it both} ejecta at 
$\sim 10\,\rm GHz$, as well as of the total fluxes at $1.4-3.3\,\rm GHz$  
are known.

\subsection{Model parameters}

For  calculation of the fluxes expected in the observer frame
we approximate the principal model parameters  as the following functions 
of energy $\gamma$ and time $t^{\prime}$ in the {\it rest frame} of 
spherically expanding cloud: 

\vspace{2mm}
\noindent
(i)~{\it Speed of expansion} is taken as
\begin{equation}
v_{\rm exp}(t^\prime) = \frac{v_0}{(1+t^{\prime}/t_{\rm exp})^k}\; \cdot
\end{equation}
For $k > 0$ this form of  $v_{\rm exp}(t^{\prime)}$  
enables deceleration of the 
cloud's expansion at times $t^{\prime} \geq t_{\rm exp}$. 
The radius of the cloud is 
\begin{equation}
R(t^\prime) = \frac{v_0 \, t_{\rm exp}}{1- k}
\left[ \left( 1+\frac{t^{\prime}}{t_{\rm exp}}\right)^{1-k}\, - \,1\,
\right]\; \cdot
\end{equation}
For numerical calculations we approximate the profile of 
$v_{\rm exp}(t^\prime)$ with
the mean $v_{\rm i}$ in the succession of short time intervals $t_{\rm i+1}-
t_{\rm i}\ll t_{\rm i}$, as explained above.

\noindent
(ii)~~The mean {\it magnetic field} is supposed to be decreasing with the 
radius $R$ as
\begin{equation}
B(t^\prime)\, = \, B_{0}\, (R/R_{0})^{-m}
\end{equation}
 Here $R_0 $ is the cloud's radius at the intrinsic time 
$t_{0}^{\prime}= 2.7\,\rm days$ after ejection (March 19.8).
This time in the rest frame of the approaching cloud corresponds 
to the observer's time $t_0=t_{0}^{\prime}/\delta = 4.8\,\rm days$ 
when the flare was
detected by the VLA telescope (March 24.6).  

\noindent
(iii)~~The {\it injection rate} of the electrons into each of the twin clouds  
is taken in the 
form of $Q(\gamma,t^{\prime}) = \,Q_{0}(\gamma) \, q(t^{\prime})$, with  
\begin{equation}
q(t^{\prime})= (1+t^{\prime}/t_{\rm inj})^{-p} \; ,
\end{equation}
which enables the decline of injection rate at 
$t^{\prime} \geq t_{\rm inj}$, and 
\begin{equation}
Q_0(\gamma) \propto \gamma^{-\alinj} e^{-\gamma/\gamma_{\rm c}} \;.
\end{equation}
In this section we do not specify the exponential cut-off energy 
$\gamma_{\rm c}$, assuming only that $\gamma_{\rm c}\geq 10^4$ to account for
the radio data. 
The coefficient of proportionality in equation (36) is chosen  
so as to provide the total flux  $655\,\rm mJy$ detected at 
$8.42\,\rm GHz$ on March 24.6 (MR94).

(iiii)~~The {\it escape time} $\tau$  is approximated after
following considerations. Let $\lambda_{\rm sc}$
is the mean scattering length of the electrons in the cloud. Then the
diffusion coefficient ${\sf D} \simeq \lambda_{\rm sc} c/3$, and
the characteristic escape time can be estimated as
$\tau \sim R^2/2 {\sf D} \simeq 1.5 R^2 /c \lambda_{\rm sc}$. If the diffusion
would be close to the Bohm limit, which corresponds to $\lambda_{\rm sc}$ 
about 
of the Larmor radius $r_{\rm L}=m_{\rm e} c^2 \gamma /e B$ of particles, then 
$\tau \propto R^{2-m}\gamma$. In this case, however, the 
escape time of $\gamma \leq 10^3$ electrons from the radio cloud
with $R_0\sim 10^{15}\,\rm cm$ and $B_0\sim 0.1 \,\rm G$ would exceed 
$10^{6} \,\rm yr$. Thus, in order to provide escape times of order of a day, 
the diffusion of GeV electrons  in the clouds should proceed many orders of
 magnitude faster than in the Bohm limit. Assuming that  
the scattering of the electrons is due to plasma 
turbulent waves with  energy density $w_{\rm turb}$, the 
frequency of collisions, $c/\lambda_{\rm sc}\,$, would be 
$\propto w_{\rm turb}$. A widely used approximation  
$w_{\rm turb}\propto B^2/8\pi$ results in  
$\tau \propto R^{2(1-m)}$. However, in principle $w_{\rm turb}$ may drop 
faster than the energy density of the magnetic field. Therefore it
seems reasonable to approximate $\lambda_{\rm sc}$ as 
\begin{equation}
\lambda_{\rm sc}/R = C_{\lambda}\,  (R/R_0)^{u} \, 
(\gamma/\gamma_\ast)^\Delta \;,
\end{equation}
with the index $u$ being a free parameter.
Then the relation $u \simeq 2m-1$ would correspond to the rate of decrease 
$w_{\rm turb}\propto w_{\rm B}$, while larger
values of $u$ could indicate a faster decay of the turbulence. 
For the normalization energy we  take $\gamma_{\ast} =2\times
10^3$ (i.e. $E_{\ast}=1 \,\rm GeV$). 
Since $\tau$ cannot be less than the time of 
rectilinear escape of electrons from the cloud, we come to   
\begin{equation}
 \tau(\gamma,R)=\frac{R}{c} \left[ 1 + \frac{3}{2 C_{\lambda}}
\left( \frac{R}{R_0}\right)^{-u} \left( 
\frac{\gamma}{\gamma_\ast}\right)^{-\Delta} \right] \; \cdot
 \end{equation}

For estimation of $C_{\lambda}$, we remind that at 
times of flare maximum $t^\prime\simeq t_0$ the electron 
distribution $N(\gamma,t^\prime)$ in the range of $\gamma\sim 300$
should be {\it not}   
modified by the escape, which requires 
$\tau(300,R_0) \geq t_0 \sim R_0/v_0$.
Meanwhile at $t^\prime \sim $ few times $t_0$ the steepening of the 
radio spectrum
is significant (see MR94), so  
$\tau[300,R(3t_0)] \leq 3\,t_0$. These two requirements result in a rough 
estimates  $C_{\lambda} \sim (0.3-1)$ if $t_{\rm exp }\gg t_0$, and
$C_{\lambda} \sim (0.1-1)$ if $t_{\rm exp }\leq t_0$. 
Therefore, characteristic scattering of the radio electrons   
should occur  on the lengthscales  
$\lambda_{\rm sc}\sim (0.01-1)\, R$. 
Since in the case of $\Delta \simeq 1$ the energy dependent term in 
equation (38) quickly decreases with increasing energy, for the electrons with 
$\gamma \gg \gamma_{\ast}$  the 
escape time is energy-{\it independent}, $\tau \simeq R/c$, 
so rather unusual spectra of relativistic  
electrons in the ejecta can be expected.

\vspace{2mm}
Thus, the principal parameters of our model are:
$v_0$, $t_{\rm exp}$ and $k$ for the
speed of expansion, $B_0$ and $m$ for the magnetic field,
$\alinj$, $t_{\rm inj}$ and $p$ for the electron injection spectrum,
$C_\lambda$, $u$ and $\Delta$ for the electron escape time. Some of these 
parameters, as $\alinj\simeq 2$ or $\Delta\simeq 1$, can be fixed rather   
well using the arguments discussed in Section 2. 
To define the range of freedom of other parameters, especially of those 
which describe the time evolution of the flare,  
detailed quantitative calculations are needed.

\subsubsection{Initial speed of expansion}
 
To show the range of possible variations of $v_0$, in Fig.~1 we present the
total fluxes of synchrotron radiation which could be expected in the 
observer frame at $t_0 =4.8\,\rm days$ after 
ejection of the radio clouds, calculated for 3 different expansion speeds 
$v_0$, and assuming 2 different magnetic fields $B_0$. 
It is seen that in the case of $B_0 = 0.2 \,\rm G$, when
magnetic field and relativistic electron energy densities are on the
equipartition level, $w_{\rm B} \simeq w_{\rm e}$, the values of 
$v_0 \simeq (0.1-0.2)\,c\,$ can explain the observed fluxes.
Assumption of $B_0 = 0.05\,\rm G$ results in a better agreement of 
calculated and observed 
fluxes for $v_0 =0.1\,c$ (dashed curves), but the values of $v_0\sim 0.05\,c$ 
 still are not  acceptable. 

In order to show that such high expansion speed of radio clouds is  
a common feature for the flares of GRS~1915+105, in Fig.~2 we present 
evolution of the fluxes calculated for the flare  
observed on July 8-10 (Gerard 1996), when 
between July 9 and July 10 apparent transition of the spectra 
 in $1.4-3.3\,\rm GHz$ band from optically thick to thin forms has occured. 
All curves in Fig.~2 are normalized
to the flux $94\,\rm mJy$ at $3.31 \,\rm GHz$ observed on July 10.0 (the time
$t=0$ in Fig.~2 corresponds to July 8.0). By appropriate choice of the  
time of ejection it is possible to explain both the  
flux $83\,\rm mJy$ detected at $3.31 \,\rm GHz$, and the flux
upper limit at $1.41 \,\rm GHz$ on July 9.0\,. 
However, the flux $140\,\rm mJy$ detected on 
July 10 at $1.41\,\rm GHz$ can be explained only assuming high speed of 
expansion of radio clouds. As it follows from Fig.~2, for $v_0 
= 0.2\,c$ and $0.15\,c$ there is a good agreement between the observed and 
calculated fluxes, whereas in the case of $v_0 = 0.1\,c$ the calculated flux 
$S_{\rm 1.4\,GHz} \simeq 105 \,\rm mJy$ is significantly below of the
$\pm 20\,\rm mJy$ flux error limits for the Nancay measurements (see 
Rodriguez et al. 1995). In Fig.~2  the magnetic field
at the time $t_{0}=1\,\rm day$  
is  $B_0=0.35\,\rm G$, which corresponds to the equipartition field for
this flare.
Decreasing $B_0$ down to $\sim 0.02\,\rm G$, when $w_{\rm e}/w_{\rm B}\geq
10^5$ (!), it is possible to fit  
the measured fluxes with $v_0 \simeq 0.1\,c$, but not significantly
less than that. 
Thus, even for relatively weak flares  one has to suppose
 very high speeds of expansion of radio clouds, 
$v_0 \simeq (0.1-0.2)\, c$.

\subsubsection{Magnetic field}

Accurate numerical calculations confirm the estimates in Section 2 for the 
equipartition magnetic field $B_{\rm eq} \sim (0.2-0.3)\,\rm G$ at 
times of flare maximum,  as well as the fact of fast increase of 
the ratio $\eta =w_{\rm e}/w_{\rm B}$ 
and required initial injection power $P_{\rm inj} = m_{\rm e} c^2 \,
\int \gamma \,Q_{0}(\gamma) {\rm d} \gamma$  of relativistic electrons 
in the case of $B_0 < B_{\rm eq}$. For example,  
the fluxes corresponding to the case of $v_0= 0.2\,c$ in Fig.~1 
require $P_{\rm inj} =1.2 \times 10^{38}\,\rm erg/s$ 
(in the energy range $\gamma \leq 10^4$) if 
$B_0=0.2\,\rm G$, and $P_{\rm inj} =10^{39}\,\rm erg/s$ if $B_0=0.05\,\rm G$. 
Note that for a fixed $B_0\,$, but different values of other model parameters 
the values of $P_{\rm inj}$ may 
change, within factor of $\sim 3$, due to different rates of adiabatic 
and escape losses. Thus, for the fluxes shown in Fig.~1 in the case of 
$B_0=0.2\,\rm G$, but $v_0= 0.1\,c$, the injection power 
$P_{\rm inj}=  4.9 \times 10^{37}\,\rm erg/s$ is needed.

In Fig.~3 we show the temporal evolution of the fluxes calculated for a 
fixed $B_0=0.2\,\rm G$ but 3 different values of the power-law index 
$m$ in equation (34), and assuming that electron escape is negligible.
One might conclude from Fig.~3 that for proper interpretation
of the observed decline of $S_{\nu}(t)$ the index $m$ is to be about $1.2-1.3$.
However, the upper panel in Fig.~3 clearly shows that without energy-dependent
escape there is no spectral steepening of the fluxes at later stages of the
flare, as observed. The dotted curve in Fig.~3, which is calculated for $m=1$ 
assuming energy-dependent escape, shows noticeable, 
but still insufficient, evolution of the spectral index $\alr$.
Simultaneously, the decline of radio fluxes becomes faster, and now  
well agrees with the observed slope of $S_{\nu}(t)$. Thus, most probably,
the decrease of magnetic field in the expanding clouds occurs with the 
power-law index $m\simeq 1$. Interestingly, for this value of $m$ one could 
expect that during the cloud's expansion the ratio  
$\eta=w_{\rm e}/w_{\rm B}$ remains at some fixed level, $\eta(t)\sim
\,\rm const$. Indeed, the magnetic field energy density in a cloud, 
expanding with some mean $\overline{v_{\rm exp}}$, is decreasing 
$\propto t^{-2m}$, while the 
time behavior of the energy density of relativistic electrons 
can be roughly estimated as $w_{\rm e} \propto \overline{q}\, t/R^3
\propto t^{-2}$, so $\eta \propto t^{2(m-1)}\sim \,\rm const$. A slow decline
of $\eta(t)$ could be expected due to escape of the electrons.
Numerical calculations support  this conclusion.   

\subsubsection{Time profiles of the expansion/injection and escape rates}

Although the dotted curve in Fig.~3 reveals noticeable steepening of the 
radio spectrum, it does not explain the 
spectral index $\alr\simeq 0.84$ measured on April 16 (MR94). 
Our attempts to fit the observed spectra with a constant speed of
cloud's expansion,  $v_{\rm exp}(t)=\,\rm const$, failed. 
As we discussed in Section 3.3, the energy
dependent escape of relativistic electrons may result in significant steepening
of $N(\gamma,t)$ only on timescales $t\gg \tau(\gamma,t)$. Meanwhile,
the spectral index $\alr\sim 0.5$ on March 24 implies $ \tau(\gamma,t)\geq t$
at $t\simeq t_0$. Then, for the escape time given by equation (38), it is 
difficult to expect sufficiently fast increase of the ratio $t/\tau$ on 
timescales of few times $t_0$, if the cloud would expand with a constant 
speed.
 
In order to enable a faster escape of the electrons, one has to 
assume that expansion of the clouds at times $t\geq t_0$ significantly
decelerates, which implies $t_{\rm exp}\leq t_{0}^{\prime}$. 
Simultaneously, the injection rate of relativistic electrons should also 
decrease, since otherwise due to a slower drop of the magnetic field in time 
(for decelerated expansion) 
the decline of $S_{\nu}(t)$ would be significantly slower than observed.
The study of possible correlations of the model parameters in Eqs.~(32) and 
(35) which could explain the evolution of both $S_{\nu}(t)$ and
$\alr(t)$, reveals that the best fits are reached when 
$t_{\rm exp}\sim$ few days, $k\sim (0.7 -1)$, and
more importantly, when the time profiles of 
the injection rate and the speed of expansion are connected with 
equations $p\simeq 2\, k$ and 
$t_{\rm inj} \sim (1-2)\, t_{\rm exp}\,$.
In Fig.~4  we show the spectra calculated for 
$t_{\rm exp}=2 \,\rm days$ and $k=0.75$, assuming different combinations
of the ratios $t_{\rm inj}/t_{\rm exp}$ and $p/k$. Note that for both
curves corresponding to $p = 2k$ (solid and dotted lines) it is possible 
to reach a better agreement with the observed slope of $S_{\nu}$ 
by a better choice of $C_{\lambda}$ or $u$. Meanwhile,
in the cases of
$p=k$ and $p = 3k$ (dashed and dot-dashed curves) the discrepancy between 
the calculated and measured fluxes may be reduced only if one 
assumes $t_{\rm inj}\ll t_{\rm exp }$ for $p=k$, or 
$t_{\rm inj}\gg  t_{\rm exp }$ for $p=3k$ . Actually, it would 
correspond to simulation of the relation $p\simeq 2k$ 
by means of changing the ratio between $t_{\rm inj}$ and $t_{\rm exp }$
for $q(t)$ and $v_{\rm exp}(t)$.

For the escape rate, the  calculations show that generally
the value of parameter $C_{\lambda}$ in equation (38) may vary in the range
of $(0.1-1)$, and
the power-law index $u$ should be $\sim 1.5$, 
which is somewhat larger than $2m-1 \simeq 1$. As discussed above, it
implies that the turbulent energy density in the expanding ejecta decreases 
faster than $w_{\rm B}$.

Thus, the model parameters can be fixed as follows:
$\alinj \approx  2$, $\Delta \approx 1$, $m \approx 1$, the expansion speed
$v_0\simeq (0.1-0.3)\, c$, depending on $B_0\simeq (0.03-0.3)\,\rm G$ at 
the time $t=t_0$, the characteristic expansion time
$t_{\rm exp}\sim$ few days (in the cloud's rest frame), 
$t_{\rm inj } \sim 1.5\, t_{\rm exp}$,
and $p\simeq 2k$ with $k\simeq (0.7-1)\,$.

\subsection{Injection powered by a beam in the jet region\,? }
 
The relation $p\simeq 2k$, with $t_{\rm inj } \sim 1.5 \, t_{\rm exp}$,
implies an interesting interpretation for the primary source of power for 
continuous injection/acceleration of relativistic electrons. 
Namely, this kind of temporal behavior of $q(t^\prime)$ corresponds,
in particular, to a  scenario when
the power $p_{\rm inj}$ of relativistic electrons injected into the cloud per 
unit surface area {\it decreases} with time as 
$p_{\rm inj}\propto (t{^\prime})^{-2}$. 
Then at initial stages, when the cloud expands with a constant speed 
$v_{\rm exp}(t^\prime)\simeq v_0$, the injection rate 
$q(t^\prime) \propto R^2 p_{\rm inj}
= \,\rm const$. At times $t^\prime \sim t_{\rm exp}$, when the expansion 
starts to decelerate, $q(t^\prime)$ also starts to decline. 
As follows from equation (33), at times $t^{\prime}\gg t_{\rm exp }$ the 
radius $R(t^\prime)\propto (t^\prime/t_{\rm exp})^{1-k}$ for $k<1$, and 
$R(t^\prime)\propto ln(t^\prime/t_{\rm exp})\sim \,\rm const$ for $k=1$.
Therefore, in the case of $k\leq 1$, the total injection power at
$t^\prime \gg t_{\rm exp}$ will be 
$q(t^\prime)\propto (t^\prime/t_{\rm exp})^{-2k}$, i.e. $p=2k$ as needed.
Interestingly,
it is possible to understand also, why $t_{\rm inj }$ in equation (35)
is somewhat larger than $t_{\rm exp}$: an expanding cloud needs 
some time
of order of $\Delta t^{\prime} \sim t_{\rm exp }$ {\it after} the beginning of 
deceleration for a  
significant deviation of $R(t^\prime)$ from the linear behavior, 
$v_0 \, t^\prime \,$.
This interpretation of the time profile of  $q(t^\prime)$  suggests 
an injection function in the form 
\begin{equation}
q_{\rm b}(t^\prime)= R^{2}(t^\prime)/(v_0  t^\prime)^{2}\; .
\end{equation}

Although the discussion above related to $k\leq 1$, the time evolution
of radio fluxes observed during the March/April 1994 flare   
can be explained with the injection rate given by equation 
(39)  even for the values of $k$ somewhat larger than $1$. 
In Fig.~5 we present the fluxes calculated 
for the injection rate $q_{\rm b}(t^\prime)$ for a fixed $t_{\rm exp}$, 
assuming 3 different values of the index $k$ in the range $(0.7-1.3)$.
The values of $v_0$ are chosen so
as for different $k$ the cloud's radius $R_0$ is the same.
Then in all 3 cases the spectral indexes $\alr$ at the instant $t_0$ 
practically coincide.

Remarkably, the decline of the specific injection rate 
$p_{\rm inj}\propto t^{-2}$ suggests a scenario where the   
injection of electrons is powered by a  conically {\it beamed continuous 
flux of energy} from the central engine. Such a beam may be supposed in 
the form of relativistic wind of magnetized relativistic plasmas and/or 
electromagnetic waves (Poynting flux) propagating in the conical
jet region. 
Then the energy flux density on the surface of a cloud
departing from the central source at a constant speed $\beta$ would decrease
$\propto t^{-2}$, as long as the total energy flux propagating in the  
jet were on a constant level. In this scenario the relativistic 
electrons may be supposed either directly supplied by the beam, or they may 
be accelerated on the {\it reverse shock} front terminating the 
wind, or both of these possibilities. Here we would notice that
 the most probable scenario for explanation of relativistic electrons 
in the Crab Nebula up to $10^{15}\,\rm eV$ implies their acceleration  
on the {\it reverse} shock, at a distances $\sim 0.1\,\rm pc$, 
which terminates the ultrarelativistic magnetized wind of (predominantly) 
$e^{+}-e^{-}$ plasma produced by the 
pulsar (e.g. Kennel \& Coroniti 1984; Arons 1996). 

Note that in the framework of the scenario where the {\it bow shock} ahead of 
the ejecta is supposed to be responsible in-situ acceleration of  
radio electrons, interpretation of $p_{\rm inj} \propto t^{-2}$ law is not
so straightforward, especially if one takes into account that the ejecta
detected during the March/April flare of GRS 1915+105 did not show any 
noticeable deceleration.  However, we cannot  exclude such a possibility,
which would require a detailed study, involving a spatial profiles of 
the density and temperature distributions of the ambient gas.
Nevertheless, for convenience further on we will refer to 
equation (39) as to the case of "beam injection", and will use this injection
profile which allows us to reduce the number of free model parameters.

\subsection{Pair of radio clouds}

So far we considered the temporal evolution of the {\it total} fluxes of 
the March/April 1994 radio flare of GRS~1915+105. Below  we 
discuss what kind of information can be found in the radio data when both 
approaching and receding radio components are resolved, as it is the case for
  GRS~1915+10 (MR94),  as well as for the second of presently known 
superluminal jet microquasar GRO J1655-40 (Tingay et al. 1995; Hjellming et al. 
1995).   

\subsubsection{ Flux ratio: an asymmetry between ejecta}   

When the pair of oppositely moving radio clouds 
can be resolved, then in addition to the 
equation for the observed angular speed $\mu_{\rm a}$ of the approaching 
cloud, two more equations become available, namely, the equation for
the angular speed of receding cloud $\mu_{\rm r}$, and the equation for 
the ratio of the flux densities detected from two sources.
The first two equations are well known, and in general case, allowing for
the parameters of the ejecta to be different, for a source at a distance 
$d$ they can be written as 
\begin{equation}
\mu_{\rm a}=\frac{\beta_{\rm a}\sin \theta_{\rm a}}
{1 -\beta_{\rm a}\cos \theta_{\rm a}}\,  
\frac{c}{d} \;\; ,
\end{equation}
\begin{equation}
\mu_{\rm r}=\frac{\beta_{\rm r}\sin \theta_{\rm r}}
{1 +\beta_{\rm r}\cos \theta_{\rm r}} \,
\frac{c}{d} \;\; ,
\end{equation}
where $\beta_{\rm a,r}$ and $\theta_{\rm a,r}$ 
are the speeds and the angles of propagation of the approaching and receding 
ejecta. 
Note that for convenience in equation (41) we have substituted 
$\theta_{\rm r}\rightarrow
180^{\circ}-\theta_{\rm r}$, so that $\theta_{\rm a} = \theta_{\rm r}$ 
for the ejecta moving in 
strictly opposite directions.

The equation for the flux ratio of the counter jets being considered,  
it should be said that usually it is supposed in the form of equation (1),
where the index $j=3$ 
for the flux ratio expected from the pair of {\it discrete} 
radio clouds,  
while the value of $j=2$ gives the brightness ratio
of {\it continuous  stationary} jets,
both  
 at equal angular separations $\phi_{\rm a}=\phi_{\rm r}$ from the 
core (e.g. Lind \& Blandford 1985).  
Note, however, that equation (1) is valid only if the jets are absolutely 
{\it identical}, which implies $\beta_{\rm a}=
\beta_{\rm r}=\beta $ and $\theta_{\rm a}= \theta_{\rm r}=\theta$, as well as 
equal intrinsic luminosities $L_{\rm a}^\prime = L_{\rm r}^\prime$ of the jets. 
 In this case Eqs.~(1), (40) and (41) are not independent, and  
{\it predict} that 
$(S_{\rm a}/S_{\rm r})_{\phi} = (\mu_{\rm a}/\mu_{\rm r})^{j+\alr}$. 
Then, for the 
observed values of $\mu_{\rm a}=17\pm 0.4 \,\rm mas/day$, 
$\mu_{\rm r}=9.0\pm 0.1 \,\rm mas/day$ and $\alr = 0.84 \pm 0.04$ (MR94), 
in the case of moving radio clouds ($j=3$) one would expect the flux ratio 
$(S_{\rm a}/S_{\rm r})_{\phi} = 13.1\pm 1.7$, 
whereas $S_{\rm a}/S_{\rm r} = 8\pm 1$ is observed. 
Remaining in the framework of assumption of identical jets, this discrepancy 
between the observed and expected flux ratios of the pair of ejecta in 
GRS 1915+105 can be explained (Bodo \& Ghisellini 1995) only if the real 
speed of radio-emitting plasma ({\it fluid}) in the jets would be different 
from the speed $\beta \simeq 0.92$ of the observed 
radio {\it patterns} attributed, in that case, to propagation of the shocks
in the fluid.

Meanwhile, it is possible 
to explain the observed flux ratio in terms of  pattern speeds 
coinciding with the speeds of the radio clouds, if we  suppose that the twin  
ejecta are {\it similar}, but not absolutely identical, allowing for some 
asymmetry between them (Atoyan \& Aharonian 1997). 
 An asymmetry between
the ejecta implies validity of at least one of the inequalities 
$L_{\rm a}^\prime \neq L_{\rm r}^\prime\,$,
$\theta_{\rm a}\neq \theta_{\rm r}\,$, or 
$\beta_{\rm a} \neq \beta_{\rm r}\,$.
Using the relation $S(\nu) = \delta^{3+\alr} \, 
S^{\prime}(\nu)$ between the apparent and intrinsic fluxes 
of a relativistically moving plasmoids, instead of equation (1), we can write
the equation for the  
flux ratio of the pair of plasmoids at equal {\it intrinsic times}
\begin{equation}
\left( \frac{S_{\rm a}}{S_{\rm r}}\right)_{t^\prime} \equiv 
\frac{S_{\rm a}}{S_{\rm r}} = 
\left(\frac{\Gamma_{\rm r}}{\Gamma_{\rm a}}\right)^{3+\alr}
\left(\frac{1+\beta_{\rm r} \cos \theta}{1-\beta_{\rm a} \cos \theta}
\right)^{3+\alr} \, \frac{L_{\rm a}^\prime}{L_{\rm r}^{\prime}}\;,
\end{equation} 
where $\Gamma_{\rm a,r}= 1/\sqrt{1-\beta_{\rm a,r}^2}$ are the Lorentz factors
of bulk motion of the ejecta. 

Since a directly measurable quantity is the ratio 
$(S_{\rm a}/S_{\rm r})_{\phi}$ 
at equal angular separations $\phi_{\rm a} = \phi_{\rm r} = \phi$,
using which the flux ratio $S_{\rm a}/S_{\rm r}$ at equal intrinsic times 
$t_{\rm a}^\prime = t_{\rm r}^\prime =t^\prime $ can be deduced, 
the relation between these two flux ratios is to be found. 
In the frame of observer the ratio  of 
times corresponding to equal angular separation of the sources from the
core is $(t_{\rm a}/t_{\rm r})_\phi = \mu_{\rm r}/\mu_{\rm a}$, while the
ratio of apparent times $t_{\rm a,r}=t^{\prime}/\delta_{\rm a,r}$
corresponding to the same intrinsic time $t^{\prime}$
is $(t_{\rm a}/t_{\rm r})_{t^\prime} = \delta_{\rm r}/
\delta_{\rm a}$.
Using equations (40) and (41), the relation between the ratios
of the apparent times corresponding to equal angular separations and
equal intrinsic times is found: 
\begin{equation} 
\left( \frac{t_{\rm a}}{t_{\rm r}}\right)_\phi = \Lambda \,
\left( \frac{t_{\rm a}}{t_{\rm r}}\right)_{t^\prime}\;\; ,
\end{equation}
where
\begin{equation}
\Lambda = \frac{\Gamma_{\rm r}\, \beta_{\rm r}\, \sin \theta_{\rm r}}
{\Gamma_{\rm a}\, \beta_{\rm a}\, \sin \theta_{\rm a}}\;\; \cdot
\end{equation}
Then, for the power-law approximation of temporal evolution of the fluxes  
$S_{\rm a,r} \propto t^{-\kappa}$, 
the relation between the flux ratios at equal intrinsic
times and equal angular separations reads
\begin{equation}
\frac{S_{\rm a}}{S_{\rm r}}\, = \,
   \Lambda^{\kappa} \, 
\left( \frac{S_{\rm a}}{S_{\rm r}}\right)_{\phi} \; \cdot
\end{equation}
Thus, in the case of symmetrical jets these two ratios coincide,
as expected. But generally they are somewhat different, depending 
on the parameters of the asymmetrical ejecta, which can be 
derived from the system of equations (40)--(42) for a  given
flux ratio at equal intrinsic times $S_{\rm a}/S_{\rm r}$, which then
can be compared with the measured ratio 
$(S_{\rm a}/S_{\rm r})_{\phi}$. 

Analytical solutions to  equations (40)--(42) are found and 
discussed in detail elsewhere
(Atoyan \& Aharonian 1997). Due to strong dependence of the ratio 
$S_{\rm a}/S_{\rm r}$ on the ratio of Lorentz-factors of
the bulk motion 
$\Gamma_{\rm a}/\Gamma_{\rm r}$,  equation (42), even the 
flux ratio  $S_{\rm a}/S_{\rm r}\sim 6$
can be easily explained with a very small difference in the speeds 
of propagation of counter ejecta $\beta_{\rm a}$ and $\beta_{\rm r}$ 
(see Fig.~6). This implies that an asymmetry in the speeds of ejecta 
can be considered as the prime reason for the discrepancy between the 
measured and `expected' flux ratios in GRS 1915+105, although some 
asymmetry in the intrinsic radio luminosities $L_{\rm a}^\prime$ and 
$L_{\rm r}^\prime$, due to somewhat different content of 
relativistic electrons and/or magnetic fields in the clouds, is possible 
as well. Any significant difference in the angles of propagation 
$\theta_{\rm a} $ and $\theta_{\rm r}$ is less
probable (at least, the ejecta move in strictly 
opposite directions on the sky, see MR94). 

In Fig.~7 we show the time evolution of the fluxes which could be expected 
from the approaching ({\it south}) and receding ({\it north}) radio clouds,
in the case of 3 different flux ratios at equal intrinsic times: 
$S_{\rm a}/S_{\rm r} = 7,\; 9$ and 13. 
It is seen, that although the total fluxes  
(as previously,
normalized to the flux observed on March 24) in all cases coincide, 
the south-to-north partition of these fluxes is essentially different, and
for the flux ratio  
$S_{\rm a}/S_{\rm r} = 7$ rather good agreement of the calculated
fluxes with the ones detected from both ejecta is reached.
The parameters of the bulk motion of approaching and receding ejecta 
in that case are equal to $\beta_{\rm a}=0.926$, $\beta_{\rm r} = 0.902$
and $\theta =70.2^\circ$, resulting in the 
Doppler factors $\delta_{\rm a}=0.55$
and $\delta_{\rm r}=0.33\,$. 
From equation (45) follows 
(for the index $\kappa \simeq 1.3$ deduced from observations, MR94)
that the flux ratio $S_{\rm a}/S_{\rm r} = 7$ at equal intrinsic times 
of the clouds corresponds to the flux ratio at equal angular separations  
$(S_{\rm a}/S_{\rm r})_{\phi} = 8.6$, which is in agreement with the 
measured flux 
ratio $(S_{\rm a}/S_{\rm r})_{\phi} = 8 \pm 1$.

\subsubsection{Synchronous afterimpulses far away from the core\,?} 

It is seen from Fig.~7 that all data points for the {\it total flux} of 
the ejecta, except for the last one (on April 30), deviate from 
the calculated total flux no more than $10\,\%$, which is not too far from
the reported accuracy  $\sim \! 5\,\%$ of the fluxes
measured by the VLA telescopes (MR94). The excess $\sim \! 25\,\%$ of the total 
flux on April 30 could be also explained, if one takes into account the
second ejection event occurred around April 23 (see MR94). 
Meanwhile, when 
we separate the fluxes measured from the {\it south} and {\it north} ejecta, 
the agreement becomes much worse. In particular, this relates
to the fluxes of both components on April 9 (corresponding to 
$t=21\,\rm days$), and to the flux of receding component on April 16 (i.e.
$t=28\,\rm days$). Changing a little the model parameters, e.g. assuming
a faster escape of the electrons, it is possible to reach a better agreement
with the flux measured from the receding cloud on April 9. However, the 
discrepancy $\sim \! (30-40)\,\%$ for the two other data points cannot be 
reduced in that way, so another explanation of these fluxes is needed. 

We would like to believe that this discrepancy is connected with a 
significant increase of the fluxes between April 4 and 5 
detected by the Nancay telescope.
It is seen from Table 1 in Rodriguez et al (1995), that
during 24 hours between these 2 days, corresponding to 
$t=16-17\,\rm days$ after the ejection event,  the fluxes at
both $1.41\,\rm GHz$ and $3.3 \,\rm GHz$ have
suddenly {\it increased} by
$\sim 30\,\%$. Although the VLA was not observing GRS~1915+105
on that days, the `echo' of that increase could be present in the flux detected
by VLA from the approaching component on the next observation date, on April 9.
The flux from the receding component being considered, one should
expect a significant {\it delay} between the times of 
observations of that event
from the {\it south} and {\it north} radio clouds, if this increase 
was due to a powerful bidirectional 
afterimpulse  from the central source which would have reached the two
clouds at equal intrinsic times $t_{\rm a}^{\prime}= t_{\rm r}^{\prime}
=t^\prime$, and therefore, at different apparent times $t_{\rm a}=t^\prime/
\delta_{\rm a}$ and $t_{\rm r}=t^\prime/\delta_{\rm r}$. 
In Fig.~8 we show the fluxes calculated  for practically the same model
parameters as in Fig.~7, but assuming that there was an additional short 
impulse of injection of relativistic electrons into both clouds
during $\Delta t^\prime \leq 1\,\rm day$ around 
intrinsic time $t^\prime =9\,\rm days$ after ejection (solid curves). 
For the calculated Doppler factors of counter ejecta $\delta_{\rm a}=0.55$ and
$\delta_{\rm r}=0.33$, this intrinsic time corresponds to the 
apparent times $t_{\rm a}=16.4\,\rm days$ and $t_{\rm r}=27.2\,\rm days$
for the approaching and receding clouds, respectively, i.e. just between
April 4-5 and April 15-16. 
It is seen from Fig.~8 that the agreement of the calculated fluxes 
with the measured ones 
now becomes better. Note that the last two data points
(on April 30) in Fig.~8 cannot be explained by another afterimpulse, since 
they coincide in time. The excess fluxes on that day are, most probably, 
connected with the second pair of plasmoids ejected around April 23 (see MR94).

If such interpretation of the data is not just an
artifact, but corresponds to reality, implications for the physics of jets
may be very important. It would mean that both
clouds are energized by the central source being even far away from
it, at distances $\gg 10^{16}\,\rm cm$ (!), or otherwise we have to rely upon 
a mere coincidence of equal intrinsic times,
as well as amplitudes, of additional injection of relativistic particles
from the bow shocks ahead of two counter ejecta, which somehow would 
impulsively increase the rate of transformation of their kinetic 
energy to the accelerated electrons.  
Energization of the clouds from the central source implies that continuous 
relativistic flux of energy ({\it the beam}), in the form of 
relativistic wind of particles and/or electromagnetic fields, 
should propagate in the jet region. Then the injection of
relativistic electrons into the clouds could be supplied directly through 
this wind and/or the wind termination {\it reverse} shock on the {\it back} 
side of the clouds.

It should be noted, that in principle the excess of the radio fluxes on 
April 9 and April 16 from approaching and receding clouds, respectively, 
could be explained also by the synchronous increase of the magnetic fields
in both clouds by $\sim 40\,\%$ (the dotted curves in Fig.~8), but not of the 
injection rate of relativistic electrons. The principle difference between the 
`magnetic' and `electronic' afterimpulses consists in different
behavior of the spectral index $\alr$: if the injection of 
relativistic electrons proceeds smoothly, the spectral index at a given 
frequency evolves also smoothly, while  the `electronic' afterimpulse can 
result in significant hardening (variations) of the spectral index $\alr$. 
Thus, the data of multi-frequency radio monitoring of the clouds will be able
to distinguish between these two options. 

Interestingly, some indication for temporary hardening of the radio spectra 
during April 4 to April 6, i.e.
 coincident with the time of the supposed `afterimpulse', can be found 
in the data of observations of the March 19 flare at frequencies 
1.41 and 3.28 GHz (Rodrtiguez et al. 1995). This effect can be seen in 
Fig.~9 where we compare our calculations with all radio data available for 
the March 19 flare. 
For the reported accuracy of the Nancy fluxes 
$\sim 5\,\%\,$ (Rodriguez et al. 1990),
the agreement between  calculated and measured 
fluxes at low frequencies is worse than at 8.4\,GHz. At 
these frequencies some allowance is to be made for possible contribution 
(flickering) of the unresolved 
central source into the detected total flux, which, in particular,
may have significant impact on the correct determination of the power-law 
spectral index $\alr$. 
Nevertheless, it should be said that 
while at 8.4\,GHz the predicted fluxes well agree with the VLA data,
at low frequencies they agree with the measurements only  
qualitatively, being {\it systematically} higher 
(by $\sim 20\,\%$) than the fluxes measured by Nancy. 

Another important feature seen in the Nancy data, is the observed {\it 
non-smooth} decline of the flare, with noticeable
variation on timescale of days on the $S_{\nu}\, t$ plot. 
If not attributed to the 
flickering of the central source, this may indicate that perhaps the 
continuous energization of the radio clouds is not as smooth as supposed 
in our model calculations. Obviously, for interpretation of the
multiwavelength radio data with accuracy $\sim 5\,\%$,
one needs  more accurate theoretical models, which would take into account 
more complex behavior of the injection
and escape of relativistic electrons, as well as the spatial non-homogeneity 
of the radio clouds.

\section{Predictions for nonthermal radiation at high energies}

In the study of radio spectra of GRS 1915+105, 
in the previous sections we did not specify the maximum energies
of the accelerated electrons, assuming only that the power-law distribution
of the electrons injected into the expanding clouds extend to energies beyond
several GeV, to account for the radio flares of GRS 1915+105 detected. 
Meanwhile, it is not 
excluded that the relativistic electrons in the jets of microquasars are
accelerated  to much higher energies, similar to the case of jets in the 
X-ray selected BL Lacs, where the ultrarelativistic electrons 
are shown up via the synchrotron
X-rays and inverse Compton TeV \grs (e.g. Urry \& Padovani 1995; 
 Ghisellini \& Maraschi 1997). 
Therefore, after determination from the radio data the range of model 
parameters of the ejecta in GRS 1915+105, below we consider
the fluxes of the synchrotron and IC radiations which could be expected
at higher photon frequences, provided that the spectra of injected electrons
would extend beyond TeV energies.

\subsection{The synchrotron radiation beyond radio domain}

Interpretation of the IR (K-band) 
jet observed from GRS 1915+105 by Sams et al. (1995) in terms of synchrotron
radiation suggests that the injection spectrum of the electrons extends 
beyond tens of GeV. Besides, the injection
of relativistic electrons should continue with rather high rate 
during the first 
$t\simeq 10\,\rm days$ or more, which are needed for the ejecta 
to reach the distances $\gg 10^{16}\,\rm cm$ from the core as observed. 
The requirement of continuous injection is  
in agreement with the conclusions derived from radio data, however, there are
also some problems, connected with the interpretation of the absolute values
of the fluxes of the IR jet. 

The reported flux of the IR jet corresponds to the level of 
maximum radio fluxes observed during the strong flares, extrapolated to the
K-band with the power-law index $\alr \simeq 0.5$. This typically implies 
relativistic electrons with power law 
index $\ale \simeq 2$. Meanwhile, although there were
no observations in the radio band at the time of detection of the IR jet,
typically at  $t\geq 10\,\rm days$ after the ejection the radio flares are 
already at the decreasing stage, when the electron distribution in the 
GeV range is to be significantly modified,  to account for 
the steep radio spectra at this stage. This is 
demonstrated in Fig.~10 where by heavy lines we show the temporal evolution of 
the spectra of electrons in the clouds, calculated 
so as to provide the flux
655 mJy at 3.5 cm on $t_0 = 4.8\,\rm days$ after ejection, as previously.
Note that the flattening of the electron spectra above 10 GeV corresponds 
to the region where electron escape becomes energy independent, being
defined essentially by their flight time across the cloud.
In Fig.~11 we show the spectra of synchrotron 
radiation of these electrons at 3 different times $t$. 
 It is seen that at $t=10\,\rm days$
the synchrotron flux in the IR band is by an order of 
magnitude less than the flux detected   
by Sams et al (1996).  

To explain the observed high IR flux, we note that the radiation spectra shown
by solid curves in Fig.~11, correspond to the emission of relativistic 
electrons {\it inside} the clouds with the mean magnetic field $B(t)$
and radius $R(t)$. Meanwhile, the electrons {\it escaping} from the region
of high magnetic field, i.e. the cloud, do not immediately disappear, but  
rather should spend some time at distances comparable with $R$, before 
departing to large distances from the cloud. In Fig.~10 the thin lines 
correspond to the energy distribution of the electrons $N^{\prime}(\gamma)$
which have already left the cloud, 
but still are in the region $\leq 3\,R(t)$. The 
calculations are done assuming that outside of the cloud the 
electrons propagate
with characteristic radial velocity $\sim \! c/3$ spiraling along the magnetic
field lines. The thin lines in Fig.~11 show the synchrotron spectra 
of those electrons calculated for the magnetic field otside the cloud
$B^{\prime}(t) =0.5\, B(t)$. Remarkably, at later stages of the flare 
the synchrotron radiation flux due to the escaped electrons may be comparable
or even exceeding the radiation of the cloud in the IR/optical region,
while in the radio domain the radiation produced in the cloud
dominates at all timescales. 
Note that at $t=10\,\rm days$ the total nonthermal IR flux 
predicted in Fig.~11 is only by factor of $3 $ smaller than the one observed.

The results presented in Fig.~11 indicate that the radiation spectra 
calculated in the framework of more realistic,
spatially inhomogeneous, model for the entire synchrotron emission region
 could be able to explain
the steepening of the radio spectra simultaneously with the high IR fluxes 
much better than the simplified homogeneous cloud model does. 
Another possibility to increase the 
flux of synchrotron photons in the IR without contradiction with the observed 
steepening of the radio spectra, is connected with the assumption of 
somewhat harder injection spectrum of relativistic electrons, 
with the power law index $\alinj \simeq (1.8-1.9)$.

If the electrons in the jets of GRS 1915+105 are accelerated beyond TeV 
energies, the contribution of the synchrotron photons may become important 
also in the X-ray/\gr domain. Most probable origin of the bulk of observed 
X-rays in this strong X-ray transient source, with the peak luminosities 
during the flares exceeding $10^{39}\,\rm erg/s$ (e.g. 
Greiner et al. 1996), is the thermal accretion plasma around the black hole.
However, since the observed X-ray spectra are rather steep, with a typical
power-law index $\alx \simeq 3$ in the region of tens of keV 
(Harmon et al. 1997), the hard synchrotron radiation 
of the jets may show up at higher photon energies.
It is seen in Fig.~11 that during the first several hours after ejection, 
when the cloud is still opaque
for synchrotron radiation at GHz frequencies, the synchrotron fluxes 
at energies $\geq 100\,\rm keV$ may dominate over the extrapolation
of thermal component. This may result in a significant flattening of the 
overall spectrum at $E \sim 100\,\rm keV$. 
Although an existence of such feature in the spectrum of GRS 1915+105 could be
seen only marginally (e.g. see Sazonov et al. 1994), in the case of
the second microquasar, GRO J1655-40, the X-ray spectra clearly extend
up to several hundreds of keV (see Harmon et al. 1995).
Interestingly,  
hard tails of the X-ray spectra are a characteristic feature  
of other representatives of the population of galactic BH candidates as well, 
such as Cyg X-1 (Ling et al. 1997) 
or 1E1740.7-2942 (Churazov et al. 1993; Wallyn et al. 1996). 
Extending this spectral feature to the extragalactic
jet sources, it is worth noting that recently variable hard X-ray spectrum 
well beyond  100 keV has been detected by BeppoSAX during the 
recent strong flare of the BL Lac source Mrk 501 (Pian et al. 1997).

\subsection{Inverse Compton gamma-rays}

Straightforward evidence for acceleration of relativistic electrons beyond
TeV energies in the jets of GRS 1915+105 could be provided by detection of 
the IC \grs at very high energies (VHE),  
$E_{\gamma}\geq 100 \,\rm GeV$. Calculations in the framework of 
synchrotron-self-Compton model show that during the strong flares one may
expect detectable fluxes of VHE $\gamma$-rays,
if the magnetic field in the ejected plasmoids would be significantly 
below of the equipartition level. In Fig.~12 we show the spectra of 
the synchrotron and IC radiations expected from GRS 1915+105 at 
$t=0.1 \,\rm day$ after the ejection event, calculated for the same model 
parameters as in
Fig.~11, except for somewhat higher exponential cut-off energy, 
$E_{\rm c} = 20 \,\rm TeV$, assuming 3 different magnetic fields 
$B_0$ at the  instant 
$t_0$. 
It is seen that for the magnetic field $B_0=0.2\,\rm G$, when the ratio of
the electron to magnetic energy densities, $\eta=w_{\rm e}/w_{\rm B}$, is
at the level close to the equipartition (see Fig.~13), the fluxes of 
the IC \grs are rather small, whereas assumption of the fields 
$B_0\leq 0.1\,\rm G$ results in a significant increase of the IC \grs.

Such a strong dependence of the expected IC \gr fluxes on the magnetic field
is explained by strong dependence of the synchrotron
radiation flux on the magnetic field,  $S_{\nu} \propto B^{1+\alr}$
(e.g., Ginzburg 1979). For magnetic fields smaller by 
factor of 2, one actually requires an increase of the injection rate of the 
electrons by factor of $\sim 3 $ to provide the same (observed) radio flux. 
Therefore, assumption of different
magnetic fields $B_0$ corresponds to assumption of different injection  
power of relativistic electrons, as shown in Fig.~13. 
The change of the magnetic field $B_0$ has 
an additional strong impact on the intensity of IC $\gamma$-rays, 
since for a given field of the soft photons
(normalized to the observed radioflux),  the increase of the 
magnetic field by factor of $a$ results in the decrease of the ratio of
the photon to magnetic field energy densities by factor of $a^2$. Therefore,
the share of the injection power of VHE electrons which is channeled 
into the IC \grs, is essentially reduced.
  
In Fig.~14 we show the time evolution of the integral fluxes of the IC \grs 
calculated for the same parameters as in Fig.~11, but for 2 different
energies of the exponential cut-off in the spectrum of injected electrons:
$E_{\rm c} = 20\,\rm TeV$ (solid curves), and $E_{\rm c} = 1\,\rm TeV$ 
(dashed curves).  The spectra of the hard X-rays/soft \grs due to synchrotron
radiation of electrons in the case of $E_{\rm c} = 20\,\rm TeV$ are also
shown (dot-dashed curves). The supposed magnetic field $B_0=0.05\,\rm G$ 
requires the `beam injection' power of the electrons $P_{\rm inj}\simeq  
4\times 10^{39} \,\rm erg/s$. Fig.~14 shows that in the case of acceleration 
of the electrons in the ejecta beyond 10 TeV, the synchrotron photons 
during the flare dominate up to $10\,\rm MeV$   
over the flux of IC $\gamma$-rays. The flux of VHE \grs may be on the 
level of the Crab flux during several hours after ejection of the radio
clouds.
In a few days the flux drops to the level of 0.1 Crab, which is still 
detectable by current Imaging Cherenkov telescopes in the Northern hemisphere
(CAT, HEGRA, Whipple), but afterwards  the source becomes invisible. 

Note that this conclusion does not principally change, even if we take into 
account the contribution due to IC scattering of the electrons after 
their escape from the cloud (shown in Fig.~14 by thin lines, and calculated 
under the same assumptions as for Fig.~11), 
which at these times may become comparable with the flux 
produced inside of the clouds. The fast drop of the flux of IC \grs  is 
connected with the decrease of the total flux of the background photons, 
and fast expansion of the clouds, both 
resulting in a drastic decrease of the density
 of target photons (most importantly, from the radio to submillimeter 
domains). Note that contribution due to upscattering of the far IR radiation
of the dust, which is possibly present in the vicinity of  
GRS 1915+105 (Mirabel et al. 1997), remains very small even if 
the energy density of the dust radiation would be as high as the one of the 
blackbody radiation, and the radiation  
temperature be small, $T\sim 100\rm\,K$, 
which is most favorable for the IC production of the VHE $\gamma$-rays.

The fluence which could be 
expected at high and very high energies on different timescales during 
the strong outburst is plotted in Fig.~15. 
For the Imaging Cherenkov telescopes,
with the atmospheric shower collection area $\geq 10^{8}\,\rm cm^2$, one could 
expect up to 
several tens
of \grs with $E\geq 300\,\rm GeV$ during few hours
of observations (one night) on the first day of the outburst.
In the range of high energies, $E\geq 100\,\rm MeV$, the  
timescales when the fluxes remain on a high level, do not exceed several hours.
For a detector as EGRET aboard of Compton Gamma Ray Observatory, 
with effective detection area $\sim 10^3\,\rm cm^2$, 
the total number of photons expected during that time do not exceed few 
tens. 
Meanwhile, for the angular resolution of the 
EGRET $\sim (3-5)^\circ$, the number 
of background photons above 100 MeV due to diffuse background 
(Hunter et al. 1997) in the direction of GRS 1915+105 
($l=45.37^\circ,\; b=-0.2^\circ$) during the same time is by factor of 2-3
more. Continuation of the observations beyond the first day cannot
noticeably improve the statistics of the high-energy photons 
due to essential drop of the
signal at $t\geq 1\,\rm day$ after ejection (see top panel in Fig.~15). 
Thus, the EGRET could detect such flares from GRS 1915+105, 
in the best case, only marginally. However, 
the future GLAST detector, with 2 orders of magnitude higher sensitivity 
due to 
essentially larger effective area and better angular resolution of the photons
(e.g. see Bloom 1996),
will be able to detect the \gr flares correlated with (or slightly preceding) 
the radio flares from GRS 1915+105 even if the electrons in the clouds are 
accelerated only to 10 GeV energy range.

\section{Discussion}

The data of radio monitoring of the galactic 
microquasars, which provide a unique opportunity to follow the evolution of 
the fluxes from the pair of relativistic ejecta on conveniently 
short time scales, 
contain large information  on the basic parameters and  
processes in relativistic jets.
 
The temporal evolution of the radio flares in GRS 1915+105, with apparent
transition of the observed fluxes from synchrotron self-absorbed to optically 
thin spectral forms at the fast rising stage, and later on decline of the 
fluxes on 
time scales of days, strongly suggests that the flares
are produced in fastly expanding radio clouds ejected from the core (MR94; 
Rodriguez et al. 1995; Foster et al. 1996). The estimates of the 
characteristic parameters of the clouds
show that in a few days after ejection the clouds should reach the size 
$R\leq 10^{15}\,\rm cm$, 
which implies the expansion speeds $\sim
(0.1-0.2)\, c\,$ in the rest frame of the ejecta. This estimate 
is in agreement with the the expansion speed deduced by Mirabel et al. (1997)
from the observations of the twin radio plasmoids of 19 March 1994 outburst,
and implies extremely large broadening of the emission lines produced,
if any, in the clouds, which could effectively prevent their detection in 
the IR band.

Estimates of the equipartition magnetic field in the clouds  result 
in the values $B_{\rm eq}\simeq 0.2-0.3\,\rm G$ at the stage of maximum
of radio flares. The synchrotron cooling time of GeV electrons in such
magnetic fields is  orders of magnitude larger than characteristic 
timescale of days during which  
the radio spectra steepen from the power-law index $\alr\sim 0.5$ 
to $\alr \sim 1$. Thus, radiative losses cannot be responsible for this effect. 
We argue that for explanation of the observed steepening of radio spectra, 
a continuous injection of relativistic electrons  
into the radio clouds is needed, the prime reason for the fast modification
of the energy distribution of the electrons being the energy dependent escape 
(due to diffusion and/or drift) of the electrons from the clouds.
Note that simultaneous injection and escape of the electrons implies that 
these two processes proceed from different sides of the cloud, presumably, in
the axceal direction for the 
injection and predominantly transverse direction for the escape.

In the case of magnetic fields close to the equipartition level, 
$B\simeq B_{\rm eq}$, the energy $W_{\rm B}+ W_{\rm el} \sim 10^{43}\,\rm erg$
accumulated in the form of relativistic electrons and magnetic fields in the 
clouds in a few days after ejection is needed. This implies continuous 
injection of relativistic electrons with the initial power 
$P_{\rm inj}\sim 10^{38}\,\rm erg/s$. However, the injection is to be 
essentially more powerful, $P_{\rm inj}\geq 10^{39}\,\rm erg/s$, if the 
magnetic fields were smaller than the equipartition field by a
factor of $2$ or more. In that case the pressure of relativistic 
electrons would
essentially dominate over the magnetic energy density in the  
clouds, and could explain, at least qualitatively, fast 
expansion of the clouds with subrelativistic speeds $v_{\rm exp} \geq 0.1\,c$. 

Significant information about the processes in  relativistic 
ejecta in GRS 1915+105 becomes available after detailed comparison of the 
radio data of the prominent 19 March 1994 outburst (MR94) with 
the results of accurate model calculations of temporal evolution of 
the fluxes,
produced by relativistic electrons in fastly expanding magnetized clouds. 
The observed rate of decline of the radio fluxes can be explained if the
magnetic field  declines with increasing radius of the cloud as 
$B\propto R^{-m}$ with the index $m\simeq 1$ (see Fig.~3). This implies
that the total energy of the magnetic field in the cloud is increasing as  
$W_{\rm B} \propto R$, so it should be either effectively created in the cloud
(presumably, due to turbulent dynamo action) or supplied from outside.
Interestingly, a similar dependence of the magnetic field in the conical
jet, $B\propto r^{-1}$, but where $r$ is the jet 
{\it cross-section radius} (and not the cloud's radius, as in our case) 
have been found by Ghisellini, Maraschi \& Treves (1985) for the BL Lac
objects. Obviously, these two approximations result in the same behavior
of the cloud's magnetic field on time, $B\propto 1/t$, as long as the 
cloud expands with a constant speed $v_{\rm exp}(t) \sim v_0$, but 
for timescales $t > t_{\rm exp}$ the decline of $B(t)$ becomes different
(slower).

The steepening of the radio spectra from the power-law index $\alr \simeq 0.5$
on March 24 to the index $\alr \simeq 0.84$ on April 16 can be explained,
if both the injection of the electrons and expansion of the cloud would 
decelerate in a few days after ejection. For interpretation of the 
temporal evolution of both the spectral index and the fluxes of the
March 19 outburst, the model parameters in equations (32)
and (35), which describe the time profiles of the expansion speed of the
clouds, $v_{\rm exp}(t^prime)$, and 
of the injection rate of relativistic electrons, $q(t^\prime)$, 
should be connected as 
$t_{\rm inj} \sim (1-2)\,t_{\rm exp}$ and $p\sim 2\, k$, with $k\sim (0.7-1)$
(see Fig.~4). Remarkably, relations of this kind can be provided, if 
we assume that the electron injection is powered by the continuous beam of 
energy, in the form of magnetized relativistic wind of particles and/or 
electromagnetic waves, propagating in the region of conical jet. 
Fig.~5 shows that the `beam injection' profile $q_{\rm b}(t)$ can readily 
explain the temporal evolution of the total 
radio fluxes observed at $8.42\,\rm GHz$ (MR94).

The data of radio monitoring of the pair of counter ejecta in
GRS 1915+105 not only allow  
determination of both the speed $\beta\simeq 0.92$ and angle 
$\theta \simeq 70^\circ$ of propagation of the ejecta (MR94), but also 
contain principal information of a new quality, which cannot be found in the
radio data when only  the approaching jet is detected. 
Indeed, interpretation of the measured flux ratio  
$(S_{\rm a}/S_{\rm r})_\phi \sim 8$
of the pair of radio condensations 
in terms of a real motion of a pair of radio clouds implies some asymmetry 
between the intrinsic parameters of the jets in GRS 1915+105. 
Fig.~6 shows that, e.g., a small difference in the speeds of propagation 
of the ejecta would be enough to account even for a flux ratio 
 $(S_{\rm a}/S_{\rm r})_\phi \sim 6$, which is by factor of 2 smaller than
the flux ratio to be expected from identical counter ejecta. 
Asymmetrical ejection of the pair of plasmoids generally would require, 
just from the momentum conservation law, a transfer of a significant 
{\it recoil} momentum to the core of ejection, which is the third
object in the interacting system "two jets + core".
Comparison of this momentum with the integrated  momentum (absolute values) 
of the gas orbiting in the inner accretion disk, which is 
the most probable site responsible for production of relativistic ejecta 
(e.g., Blandford \& Payne 1982; Begelman et al. 1984), 
suggests (Atoyan \& Aharonian 1997) that
asymmetrical ejection of pair of plasmoids would be able to induce  
significant structural changes, or even destruction, of the inner disk,  
resulting in a temporary reduction/termination of the fuel supply into this
region responsible for the thermal X-rays.

Thus, in the framework
of this scenario, the onset of subcritical/supercritical accretion
would correspond to the active state of the source, with the X-ray flares which 
can proceed both with and without powerful ejection events and, therefore, 
observable radio flares (Foster et al. 1996; Tavani et al. 1996).
Powerful ejection of the 
radio emitting material may be accompanied by significant  destruction of the
inner accretion zone, which may lead to a strong decline of the X-ray fluxes
{\it simultaneously} with production of relativistic ejecta.
The time lag, up to few days, between the decline of the X-ray fluxes and
appearance of strong radio flares, as observed in both superluminal 
microquasars GRS 1915+105 (Foster et al. 1996; Harmon et al. 1997)
and GRO J1655-40 (Tingay et al. 1995; Harmon et al. 1995; Hjellming \&
Rupen 1995), corresponds to the time needed for
expansion of the clouds to become optically transparent with respect to
synchrotron self-absorption. 
Depending on the time needed for the recovery of the inner accretion disk
after ejection, the radio flares may appear just in the
dips between subsequent X-ray flares, as frequently observed 
(e.g. Harmon et al. 1997).

The data on the radio flux evolution of the pair of resolved ejecta 
contain important information on the mechanisms of continuous energization of 
the plasmoids. The interpretation of the $\sim 30\,\%$ excess of the fluxes 
detected at $8.42\,\rm GHz$ from the approaching and receding ejecta of 
GRS 1915+105 on April 9 and April 16, respectively, as the result of 
synchronous short-term increase ({\it `afterimpulse'}) of the injection rate 
of relativistic electrons into both clouds, as discussed in Section 4, implies 
continuous energization of the clouds by the central source. Then, the 
energy is to be supplied up to distances $\geq 3\times 10^{16}\,\rm cm$
by bidirectional beam, presumably in the form of relativistic wind of 
particles and electromagnetic fields, emerging from the core. Note that
the discussed above relations between  parameters of the injection rate and 
cloud's expansion, which can be provided  by the `beam injection' profile 
given by equation (39), 
can be regarded as 
another indication of injection of a real beam into the clouds.    
  
Confirmation of this scenario by future multiwavelength radio observations,
if GRS 1915+105 will give us an opportunity to trace another pair of powerful 
and long lived jets similar to 19 March 1994 event, would suggest
significant softening of the enormous energetical requirements rising in the
conventional relativistic jet scenario which implies the kinetic energy of 
the bulk motion of the ejecta be the energy reservoir for 
in-situ acceleration of the electrons on the bow shocks formed
ahead of the ejecta. Indeed, the total energy of the radio electrons injected
into the clouds during first few days with 
the power $P_{\rm inj} \sim 10^{39}\,\rm erg/s$  is about $W_{\rm inj} \sim
3\times 10^{44}\,\rm erg$, and this energy is by a 
factor of $3$ larger for all duration of the 19 March flare.  
If this amount of energy were due to kinetic energy of the bulk
motion, $W_{\rm kin}$, then the latter would be estimated as at least by 
another factor of 3 higher than  $W_{\rm inj}$, since there was not revealed  
 any deceleration of the bulk motion of the plasmoids beyond 
April 1994 (see MR94). Thus,  $W_{\rm kin} \geq
3\times 10^{45}\,\rm erg $ would be needed. Assuming that this amount of 
energy is transferred to the bulk motion `impulsively', during a very short 
timescale (since relativistic speeds imply that in a hundred of seconds 
the ejecta will be far away from not only the inner accretion disk, 
but the entire binary), one has to suppose enormous power for acceleration of 
the ejecta, $P_{\rm jet}\gg 10^{43}\,\rm erg/s$. 

Meanwhile, in the scenario which implies {\it continuous energy supply by the 
beam}, the beam power is to be about  $P_{\rm inj}$, i.e. 
comparable with the super-Eddington luminosities {\it observed} during
the X-ray flares (e.g. Greiner et al. 1996). Moreover, such a beam of 
relativistic energy can push the cloud forward against the ram pressure
of the external medium, therefore at initial stages the plasmoids may move even
with some acceleration, until they would acquire enough mass and kinetic 
energy to continue the flight ballistically. Depending on the parameters 
of the external medium, the stage of significant acceleration of the ejecta
could last, presumably, up to several days, which may be not enough to see
this effect. Note that for GRO J1655-40, which is essentially closer to us, 
the possibility to resolve initial acceleration, perhaps, could be better. 
Remarkably, in the beam model the acceleration of electrons may 
occur through the relativistic wind and/or the wind termination shock on the 
reverse (i.e. facing the central engine) side of the cloud , 
although a possible contribution due to 
acceleration on the bow shock ahead of the cloud should not be excluded
as well. 

Predictions for the fluxes expected beyond radio domain due to synchrotron 
and inverse Comton radiation mechanisms  essentially depend on the maximum
energy of accelerated electrons.  Assumption of the electrons accelerated 
beyond 10 GeV would be enough for explanation of the IR jet of GRS 1915+105 
observed by Sams et al (1996), and strongly variable IR flares detected 
by Fender et al (1997). Electrons with energies $\sim 1\,\rm TeV$  
produce synchrotron X-rays in the keV band, but even 
in the case of injection of relativistic electrons
with $P_{\rm inj} \sim 3\times 10^{39}\,\rm erg/s$ (this corresponds to
$B_{0}\simeq 0.05\,\rm G$), the flux 
of these photons cannot exceed  
$\sim \! 10\,\%$ of the thermal X-rays of the accretion disk. 

However, synchrotron  radiation may show up in the range of hard
X-rays/soft $\gamma$-rays, if the spectrum of electrons
extends to energies $\geq 10 \,\rm TeV$. In that case during first 
several hours (up to 1 day) after the ejection event, the hard synchrotron 
X-ray 
fluxes in the range of $E\geq 100\,\rm keV$ could significantly exceed the 
steep fluxes of thermal X-rays. In the case of 
$P_{\rm inj} > 10^{39}\,\rm erg/s$ (i.e. if the magnetic fields in the clouds
are by factor of 2-4 smaller than the equipartition  field)  the fluxes 
on the level $I(\geq 100\,\rm keV) > 10^{-3}\,\rm ph/cm^2 s$ can be expected,
which could be observed by the detectors like BATSE, SIGMA, OSSE, or  
BeppoSAX, currently 
operating in this energy range. Meanwhile, if the magnetic and electron energy
densities in the ejecta are about of equipartition, so the injection 
rate of electrons is $P_{\rm inj} \sim 10^{38}\,\rm erg/s$, 
then the expected maximum
fluxes of the synchrotron X-rays will be smaller by one order of magnitude.  
Remarkably, even these fluxes will be possible to detect with the forthcoming 
INTEGRAL mission intended for operation in the energy range 
$0.1-10\,\rm MeV$. The lack of detection of variable fluxes on the level
of $I(\geq 100\,\rm keV) \geq 10^{-4}\,\rm ph/cm^2 s$ during the first day  
after ejection event will mean that
the maximum (cut-off) energy of relativistic electrons in the ejecta is below
TeV domain.

Another, and most straightforward, evidence for TeV electrons in GRS~1915+105
could be provided by detection of VHE $\gamma$-rays. 
Calculations in the framework
of synchrotron-self-Compton model
 show that during strong flares one may expect
detectable fluxes of TeV $\gamma$-rays,  provided that the
the injection power of relativistic electrons 
$P_{\rm inj} \geq 10^{39}\,\rm erg/s$ (i.e. magnetic field in the radio clouds 
significantly below equipartition level).
The time evolution of the IC \gr fluxes shown in  Figs 14 and 15, 
predicts that  during first several hours of a strong outburst the
$\gamma$-ray fluxes above several hundred GeV
are on the level of $3\times 10^{-11}\,\rm erg/cm^2 s$, which corresponds to
the level of VHE \gr flux of the Crab Nebula. In a few days 
the flux drops to the level $0.1\,\rm Crab$ which can be still detected 
by current Imaging Cherenkov telescopes. Further on, however, the source
is not detectable. 

Assumption of a lower magnetic field would result in higher fluxes. However,
as far as a magnetic fields significantly less than $B_0 \simeq 0.05\,\rm G$ 
would imply injection power
of electrons $P_{\rm inj} \geq 10^{40} \,\rm erg/s$, it is difficult
to expect \gr fluxes essentially exceeding the ones shown in Figs 14 and 15.
On the other hand, in the case of in-situ acceleration/injection of the 
electrons with $P_{\rm inj} \sim 10^{38} \,\rm erg/s$ and 
magnetic fields on the level of equipartition
($B_0\simeq (0.2-0.3)\,\rm G$ at times when $R_0\simeq 10^{15}\,\rm
cm$), the IC fluxes dramatically decrease.
Therefore either positive detection or upper limits of VHE \gr fluxes,
being combined with hard X-ray observations above 100\,keV, could provide
robust constraints on the magnetic fields and 
efficiency of acceleration of electrons beyond TeV energies.

It is important to note, however, that due to narrow field of view
of the Cherenkov telescopes and possibility to do observations only during few
night hours per day, the detection of the episodes of VHE \gr emission
will be not an easy task. Fortunately, the predicted duration of high fluxes
of the VHE \grs is up to few days (see Fig.~15). Then, provided that our  
interpretation of the observed  
anticorrelation/delay between strong X-ray and 
radio flares given above
is the case, the instant of a powerful ejection event, 
and therefore the time (i.e. subsequent first night) 
most favorable for the VHE observations, can be 
determined from a strong temporary drop of the thermal X-ray fluxes
during the high state of the source. Importantly, occurrence of each strong 
ejection event can be checked retrospectively by appearance of a strong   
radio flare which is `delayed' from the time of ejection by 1-3 days. 

Information about the IC \grs can be obtained also in the range 
of high energies, 
$E \geq 100 \,\rm MeV$. It should be noted, however, that even 
during the first few hours of the flare, the fluxes expected in this energy 
range do not essentially exceed $10^{-6}\,\rm ph/cm^2 s$. 
This implies that in the optimistic case,  only marginal 
detection of such \gr flare can be expected for the EGRET which has 
effective detection area $\sim 10^3\,\rm cm^2$ and angular resolution 
of few/several degrees. We can predict, however, that the future GLAST 
instrument,
with 2 orders of magnitude higher sensitivity than EGRET, will be able to 
see the \gr flares which would precede, up to few days, strong radio flares 
from GRS 1915+105.  
\vspace{5mm}
  
\noindent
{\bf Acknowledgments}~~~We express our thanks to H. J. V\"olk for 
valuable discussions.
The work of AMA was supported through the Verbundforschung
Astronomie/Astrophysik of the German BMBF under the grant No. 05-2HD66A(7).

\clearpage

\centerline{\large {\bf References}}

\noindent
Arons J., 1996,  Space Sci. Rev.,  75, 235

\noindent
Atoyan A. M.,  Aharonian F. A., 1997, ApJ, 490, L149

\noindent
Begelman M. C., Blandford R. D., Rees M. J., 1984,
Rev. Mod. Phys., 56, 255

\noindent
Blandford R. D.,  Payne D. G., 1982, MNRAS, 199, 883

\noindent
Bloom E. D., 1996, Sp. Sci. Rev., 75, 109

\noindent
Bloom S. D., Marscher A. P., 1996, ApJ, 461, 657

\noindent
Bodo G., Ghisellini G., 1995, ApJ, 441, L69

\noindent
Castro-Tirado A. J., Brandt S., Lund N., Lapshov I., Sunyaev R. A.,
Shlyapnikov A. A.,
 
~~~~~~Guziy S., Pavlenko E. P., 1994, ApJS, 92, 469

\noindent
Churazov E., et al., 1993, A\&AS, 97, 173

\noindent
Fender R. P., Pooley G. G., Brocksopp C., Newell S. J., 1997, MNRAS,
in press 

~~~~~~(astro-ph/9707317)

\noindent
Foster R. S., et al., 1996, ApJ, 467, L81

\noindent
Gerard E., 1996, IAU Circ.\,6432

\noindent
Ginzburg V.L., 1979, Theoretical Physics and Astrophysics,
Pergamon, Oxford

\noindent
Ginzburg V. L., Syrovatskii S. I., 1964, Origin of Cosmic Rays, Pergamon,
London

\noindent
Ghisellini G., Maraschi L., Treves A., 1985, A\&A, 146, 204

\noindent
Ghisellini G., Maraschi L., 1997, in Miller H. R., Webb J. R., eds,
Blazar Variability, 

~~~~~~PASP (Conf. Ser.), in press 

\noindent
Greiner J., Morgan E. H., Remillard R. A., 1996, ApJ, 473, L107

\noindent
Harmon B. A., et al., 1994, The second Compton Symposium, AIP Conf. Proc.
304, p.210
 
\noindent
Harmon B. A., et al., 1995, Nature, 374, 704 

\noindent
Harmon B. A., Deal K. J., Paciesas W. S., Zhang S. N., Robinson C. R.,
Gerard E., 

~~~~~~Rodriguez L. F., Mirabel I. F., 1997, ApJ, 477, L85
 
\noindent
Hjellming R. M.,  Rupen M. P., 1995, Nature, 375, 464

\noindent
Hunter S. D., et al., 1997, ApJ, 481, 205

\noindent
Kardashev N. S. 1962, Sov. Astron. - AJ, 6, 317

\noindent
Kennel C. F.,  Coroniti F. V.,  1984, ApJ, 283, 694

\noindent
K\"onigl A., 1981, ApJ, 234, 700

\noindent
Landau L. D., Lifshitz E. M.,  1963, Electrodynamics of Continuous Media,
Pergamon, Oxford.

\noindent
Liang E., Li H., 1995, A\&A, 298, L45

\noindent
Lind K. R.,  Blandford R. D., 1985, ApJ, 295, 358
     
\noindent
Ling J. C., et al., 1997, ApJ, 484, 375.

\noindent
Marscher A. P., 1980, ApJ, 235, 386

\noindent
Meier D., 1996, ApJ, 459, 185

\noindent
Mirabel I. F.,  Rodriguez L. F. 1994, Nature, 371, 46

\noindent
Mirabel I. F.,  Rodriguez L. F. 1995,
Annals NY Academy of Science, 759, 21

\noindent
Mirabel I. F., Bandyopadhyay R., Charles P.A., Shahbaz T., 
Rodriguez L. F., 1997, 
 
~~~~~~ApJ, 477, L45 

\noindent
Pacholczyk A. G., 1970, Radio Astrophysics, Freeman, San Francisco

\noindent
Park B. T., Petrosian V., 1995, ApJ, 446, 699

\noindent
Pian E., et al., 1997, ApJ, (submitted)

\noindent
Rodriguez L. F., Gerard E., Mirabel I. F., Gomez Y., 
Velazquez A., 1995, ApJS, 101, 173.

\noindent
Sams B. J., Eckart A.,  Sunyaev R., 1996, Nature,  382, 47

\noindent
Syrovatskii S. I., 1959, Sov. Astron. -AJ, 3, 22.

\noindent
Tavani M., Fruchter A., Zhang S. N., Harmon B. A., Hjellming R. N., 
Rupen M. P., Baylin C., 

~~~~~~Livio M., 1996, ApJ, 473, L103

\noindent
Tingay S. J., et al., 1995, Nature, 374, 141

\noindent
Urry M. C., Padovani, P. 1995, PASP, 107, 803

\noindent
Wallyn P., Ling J. C., Wheaton Wm. A., Mahoney W. A., Radocinski R. G.,
Skelton R. T.,  

~~~~~~1996, A\&AS, 120, 295

\clearpage

%fig.1
\begin{figure}
\epsfxsize=13.5 cm
\epsffile[40 443 477 758]{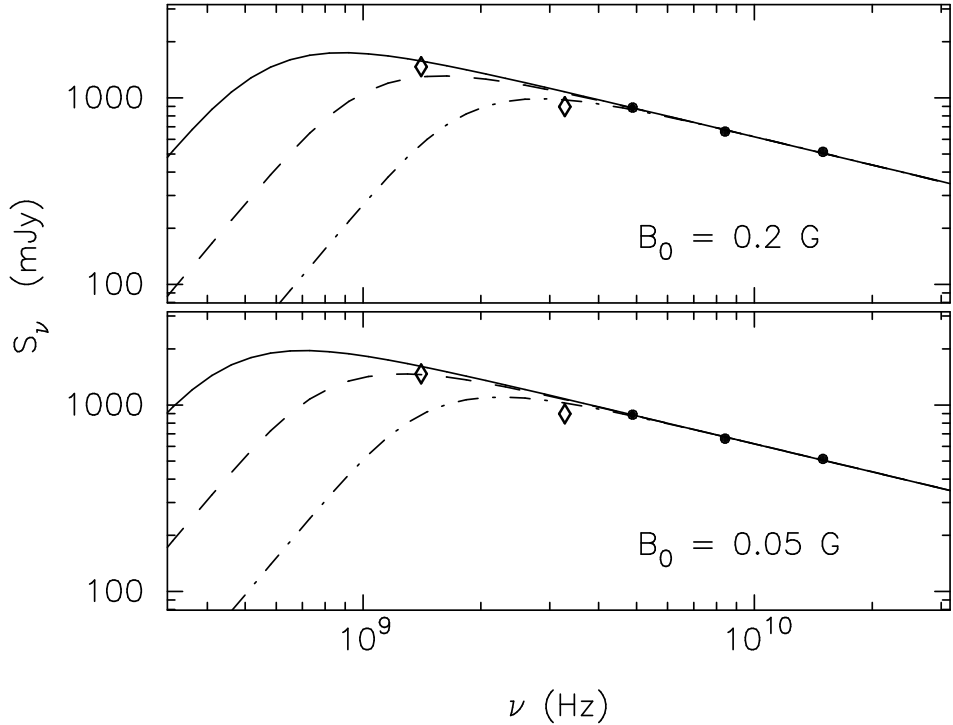}        
\caption{The spectral fluxes expected from  
GRS 1915+105 at $t_0 = 4.8 \,\rm days$ after ejection of a pair of 
relativistic plasmoids, $\beta =0.92$, in the case of 
different magnetic fields $B_0$ at the instant $t=t_0$ and 
different expansion speeds $v_{\rm exp}(t^\prime)=v_0$.
The curves shown by solid, dashed, and dot-dashed lines correspond to 
$v_0 = 0.2\,c\, , \; 0.1\,c\, ,$ and $0.05\,c\, ,$ respectively. 
Other model parameters in Eqs.\,(32)--(38) are\,: $m=1$, $\alinj = 2$, 
$t_{\rm inj}=20\,\rm days$ ($\gg t_{0}^{\prime}$), 
and $C_{\lambda}\rightarrow 0$. All curves are normalized to the flux 
$655\,\rm mJy$ at $\nu = 8.42\,\rm GHz$. The fluxes
measured by the Nancay and the VLA radio telescopes 
(Rodriguez et al. 1995) on 24 March 1994 are shown 
by diamonds and full dots, respectively. 
The $5\,\%$ error bars for the
reported accuracies of these measurements are about of 
the size of the full dots. Note that the VLA measurements
have been done $0.3\,\rm days$ after the Nancay ones. Our
calculations, however, show that the fluxes during that short time interval
change only by several per cent.}
\end{figure}

\newpage

%fig.2
\begin{figure}
\epsfxsize=13. cm
\epsffile[40 424 449 773]{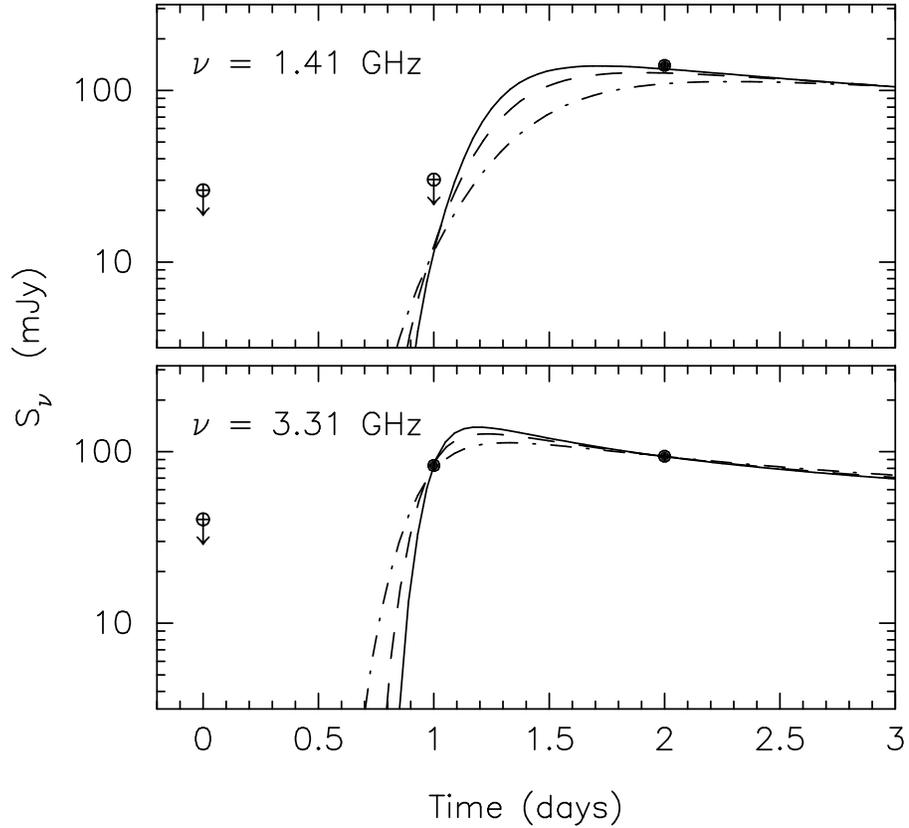}        
\caption{The temporal evolution of the fluxes at 1.41\,GHz and 3.31\,GHz for
the radio flare of GRS 1915+105 observed by Nancy telescope between July 8.0
and July 10.0, 1996 (Gerard 1996), calculated for 3 different expansion speeds:
$v_0 = 0.2\,c$ (solid curves), $0.15\,c$ (dashed curves), and $0.1\, c$ 
(dot-dashed curves). Correspondingly, 3 different ejection times -- 
July 8.8, July 8.73, and July 8.6, are supposed to fit the fluxes detected 
(full dots) at 3.31\,GHz on July 9.0 and July 10.0. 
The instant $t=0$ corresponds to 
July 8.0. The open dots show the flux upper limits  
on July 8 and July 9. }
\end{figure}

\clearpage

%fig.3
\begin{figure}
\epsfxsize=12.5 cm
\epsffile[20 28 505 540]{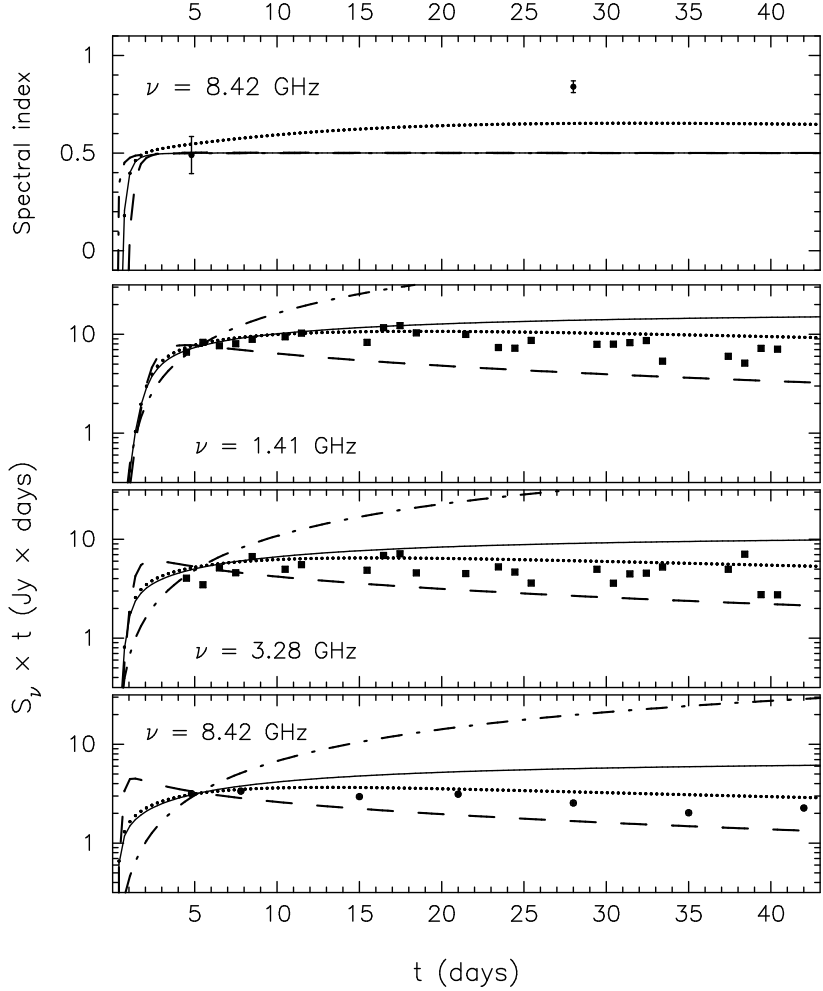}        
\caption{The temporal evolution of the fluxes at 1.41\,GHz, 3.28\,GHz 
and 8.42\,GHz, and of the spectral index at 8.42\,GHz, calculated for 
3 different values of the index 
$m$ in equation (34), assuming negligible escape of the electrons
from the cloud which expands with a constant speed $v_0=0.2\,c$. The curves
plotted by solid, dashed, and dot-dashed lines correspond to $m=1$, $m=1.5$,
and $m=0.5$, respectively. The dotted curves are calculated for $m=1$
in the case of significant energy-dependent escape of the electrons, with
parameters $\Delta =1$ and $C_{\lambda}=0.5$ in the equation (38). 
Other model parameters are the for all curves: $B_0 = 0.2\,\rm G$, 
$\alinj =2$, $t_{\rm inj}= 20\,\rm days$ and $p=1$. The  
full squares correspond to the fluxes of the 19 March 1994 flare measured 
by Nancy telescope (from Rodriguez et al. 1995), and the full dots 
show the fluxes of the VLA (MR94). $t=0$ corresponds to the supposed 
time of ejection (March 19.8).}
\end{figure}

\clearpage

%fig.4
\begin{figure}
\epsfxsize=13.5 cm
\epsffile[20 86 504 586]{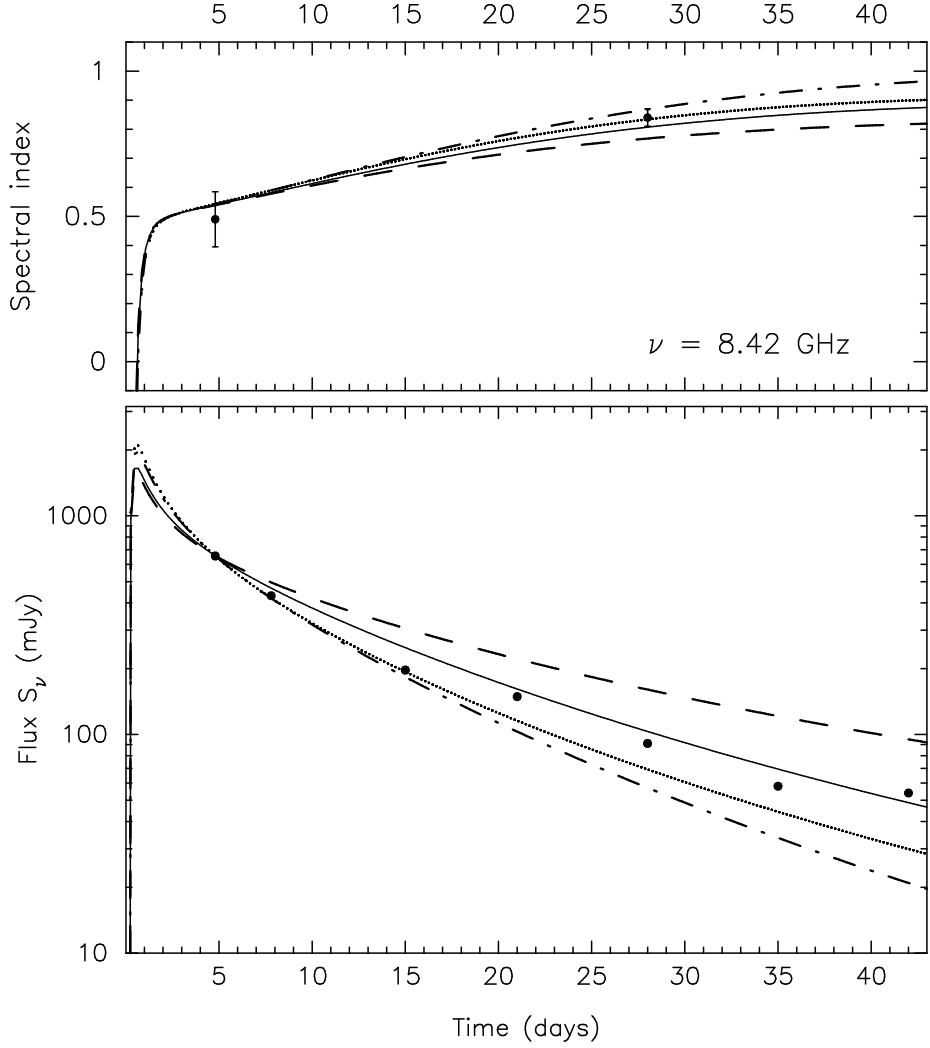}        
\caption{The evolution of the flux $S_{\nu}$ and spectral index $\alr$ 
at 8.42\,GHz expected in the case of  different relations between the 
expansion and escape model parameters in equations (32) and (35): $p= 2k$ and
$t_{\rm inj}= 2\,t_{\rm exp}$ (solid lines), $p= k$ and 
$t_{\rm inj}= t_{\rm exp}$ (dashed lines), $p= 3\,k$ and 
$t_{\rm inj}= 2\,t_{\rm exp}$ (dot-dashed lines), $p= 2k$ and 
$t_{\rm inj}= t_{\rm exp}$ (dotted lines). For all cases $k=0.75$ and
$t_{\rm exp} = 2\,\rm days$ are supposed. Other model parameters are\,:
$\alinj =2$, $v_0=0.2\,c$, $B_0=0.3\,\rm G$, $m=1$, $C_\lambda = 0.35$, 
$u=1.5$ and $\Delta = 1$.}
\end{figure}
 
\clearpage

%fig.5
\begin{figure}
\epsfxsize=13.5 cm
\epsffile[20 86 504 586]{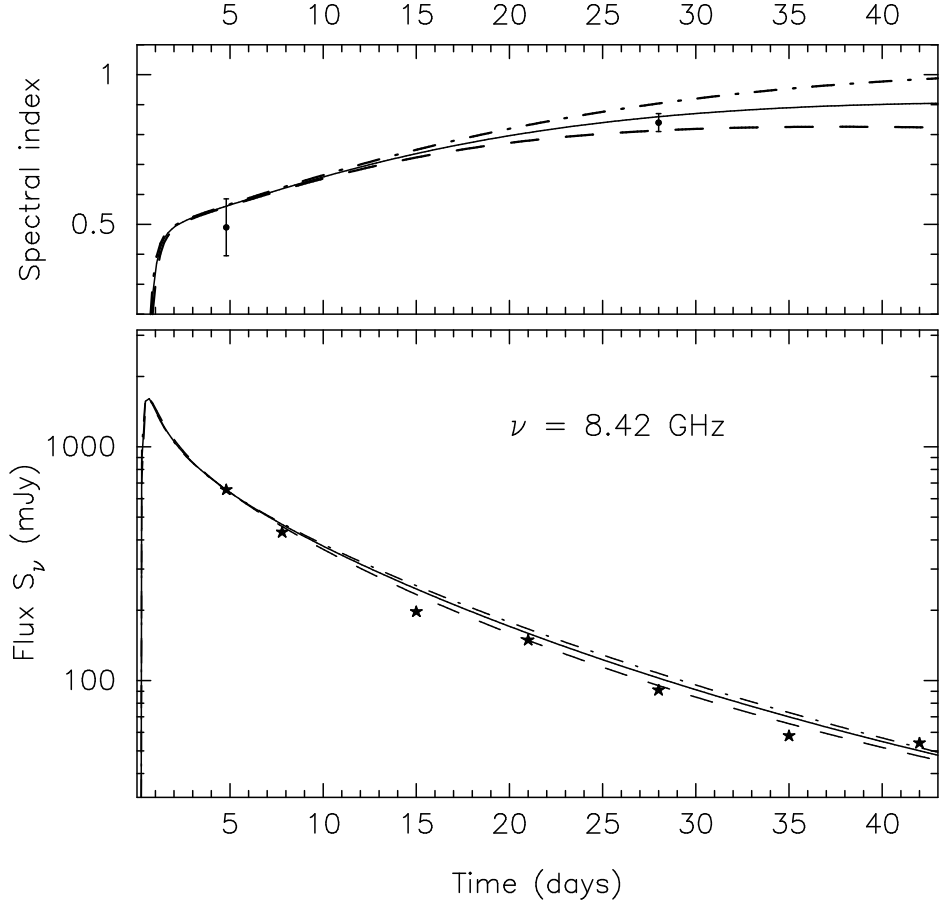}
\caption{The evolution of the fluxes and spectral indexes at 8.42\,GHz,
calculated for the `beam injection' profile of relativistic
electrons, $q_{\rm b}(t^\prime)$, 
assuming 3 different profiles for the expansion speed 
$v_{\rm exp}(t^\prime)$, with the index $k=1$ (solid curves), $k=0.7$ 
(dashed curves), and   
$k=1.3$ (dot--dashed curves). In all cases the time $t_{\rm exp} =2.5\,\rm 
days$, and the initial expansion speeds $v_0$ are chosen so as to provide 
the same radius of the clouds, $R_0=7.1\times 10^{14}\,\rm cm$,
at the instant $t_0=4.8\,\rm days$: 
$v_0= 0.15\,c$, $0.135\,c$, and $0.165\,c$, for the solid, dashed, and 
dot--dashed curves, respectively. 
 The escape parameters $C_\lambda=0.25$, $u=1.5$ and $\Delta=1$, and 
the magnetic field $B_0 = 0.1\,\rm G$ are supposed.}
\end{figure}

\clearpage

%fig.6
\begin{figure}
\epsfxsize=13.5 cm
\epsffile[30 212 513 552]{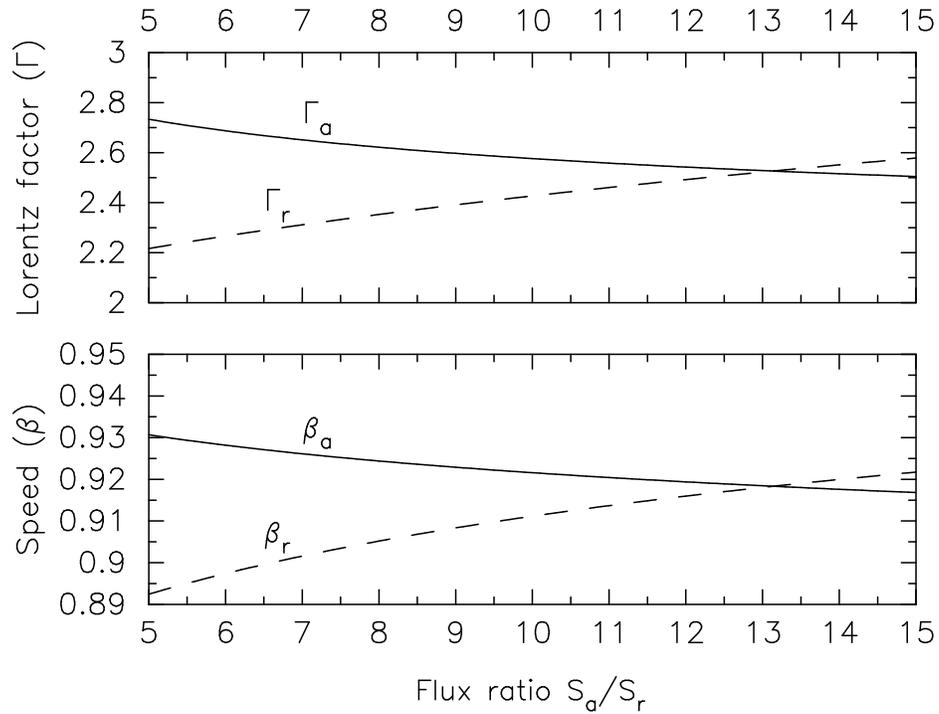}
\caption{The speeds (bottom panel) and the Lorentz factors
(top panel) of the bulk motion of the pair of ejecta calculated
for a distance $d= 12.5\,\rm kpc$, and angular
velocities $\mu_{\rm a} $ and $\mu_{\rm  r}$ observed from GRS 1915+105 
(MR94), assuming different flux ratios $S_{\rm a}/S_{\rm r}\,$.}
\end{figure}

\clearpage

%fig.7
\begin{figure}
\epsfxsize=13. cm
\epsffile[30 44 504 393]{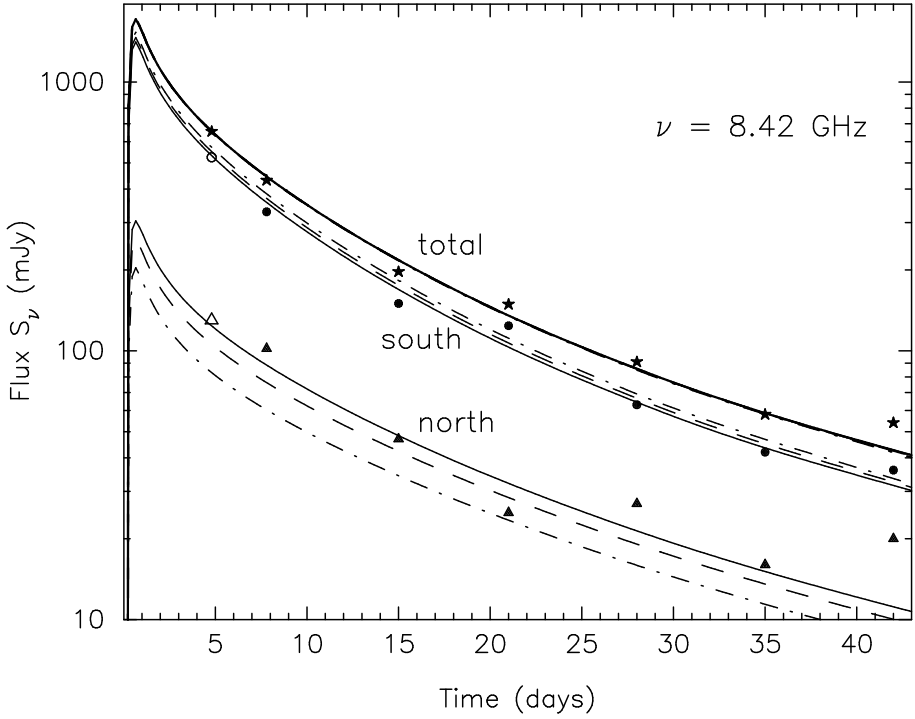}
\caption{The fluxes expected
from the approaching ({\it south}) and receding ({\it north}) radio clouds,
in the case of 3 different flux ratios at equal intrinsic times,
$S_{\rm a}/S_{\rm r} = 7$ (solid curves), 9 (dashed curve), and 13 (dot-dashed
curve), calculated under assumption of equal intrinsic luminosities, 
$L_{\rm a}^\prime = L_{\rm r}^\prime$, but allowing for the speeds of the
bulk motion of the `twin' ejecta to be different, which  
in these 3 cases are equal to 
($\beta_{\rm a}=0.926$, $\beta_{\rm r}=0.902$),
($\beta_{\rm a}=0.923$, $\beta_{\rm r}=0.908$),
and ($\beta_{\rm a}= \beta_{\rm r}=0.918$). The clouds expansion speed
$v_0=0.14\,c$ and escape parameter $C_\lambda=0.3$ are supposed. All other
model parameters are equal to the ones as in the case of $k=0.7$ in Fig.5. 
 The data points plotted by stars correspond to the total fluxes, and the full
dots and full triangles show the fluxes detected by Mirabel \& Rodriguez
(1994) from the approaching and receding components of the March 19 radio flare.
The open dot and open triangle, 
corresponding to the fluxes 525\,mJy and 130\, mJy, respectively, are 
plotted for demonstration of a possible sharing of the 655\,mJy 
total flux between southern and northern components which were not  
resolved on March 24 (see MR94).}
\end{figure} 

\clearpage

%fig.8
\begin{figure}
\epsfxsize=13. cm
\epsffile[20 80 503 587]{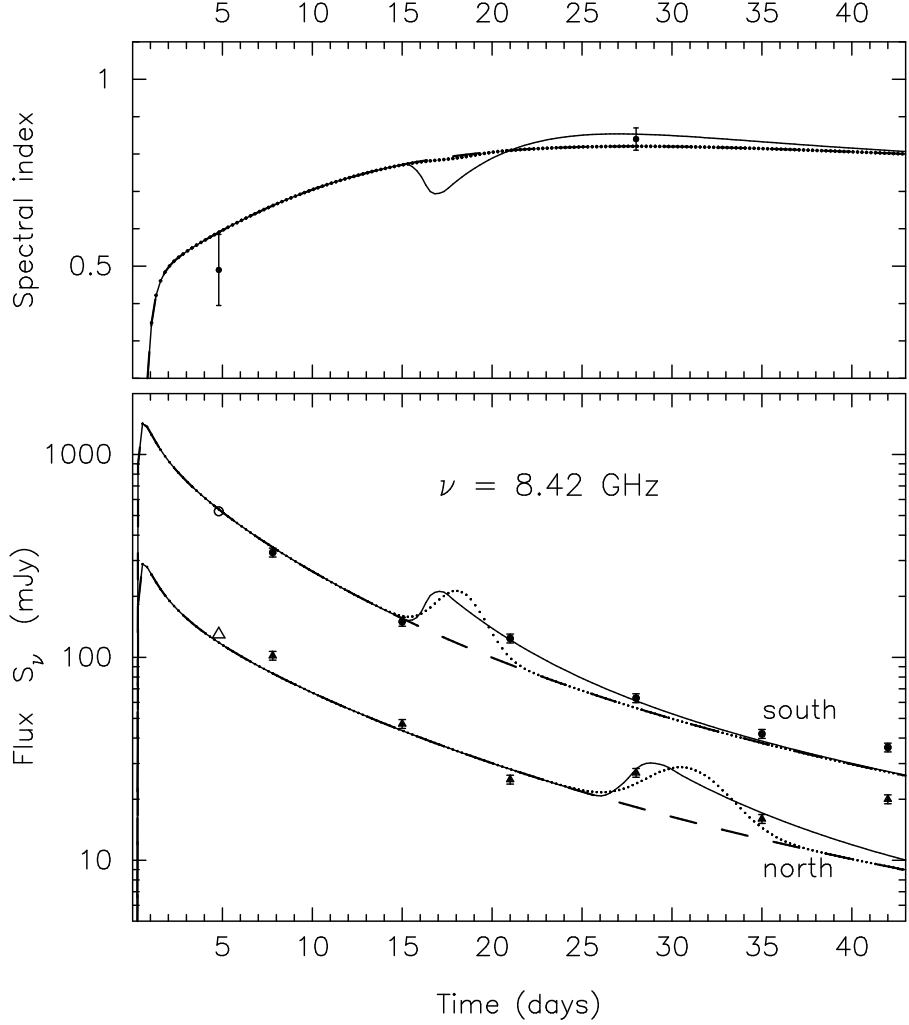}
\caption{The temporal evolution of the fluxes (bottom panel)  
expected from the southern and 
northern radio clouds of the March 19 outburst in the case of synchronous 
impulsive increase of the injection rate of relativistic electrons 
(`electronic afterimpulse', solid lines) or of the magnetic field 
(`magnetic afterimpulse', dotted lines) in both plasmoids during $\Delta t\leq
1\,\rm day$ around intrinsic time $t^\prime = 9\,\rm days$ after ejection 
event. The top panel shows the evolution of the spectral index 
at 8.42\,\rm GHz expected for the {\it southern} component.
The dashed curves correspond to the case of smooth behavior of the
electron injection and magnetic field (i.e., without any afterimpulse). 
The `beam injection',  with the same model parameters as in Fig.~7 
(for $S_{\rm a}/S_{\rm r} = 7$), is supposed, 
except for $k=1$ and $C_\lambda=0.4$.
The error bars on the bottom panel correspond to the $5\,\%$ accuracy of the
measured fluxes.} 
\end{figure}

\clearpage

%fig.9
\begin{figure}
\epsfxsize=13. cm
\epsffile[15 16 503 704]{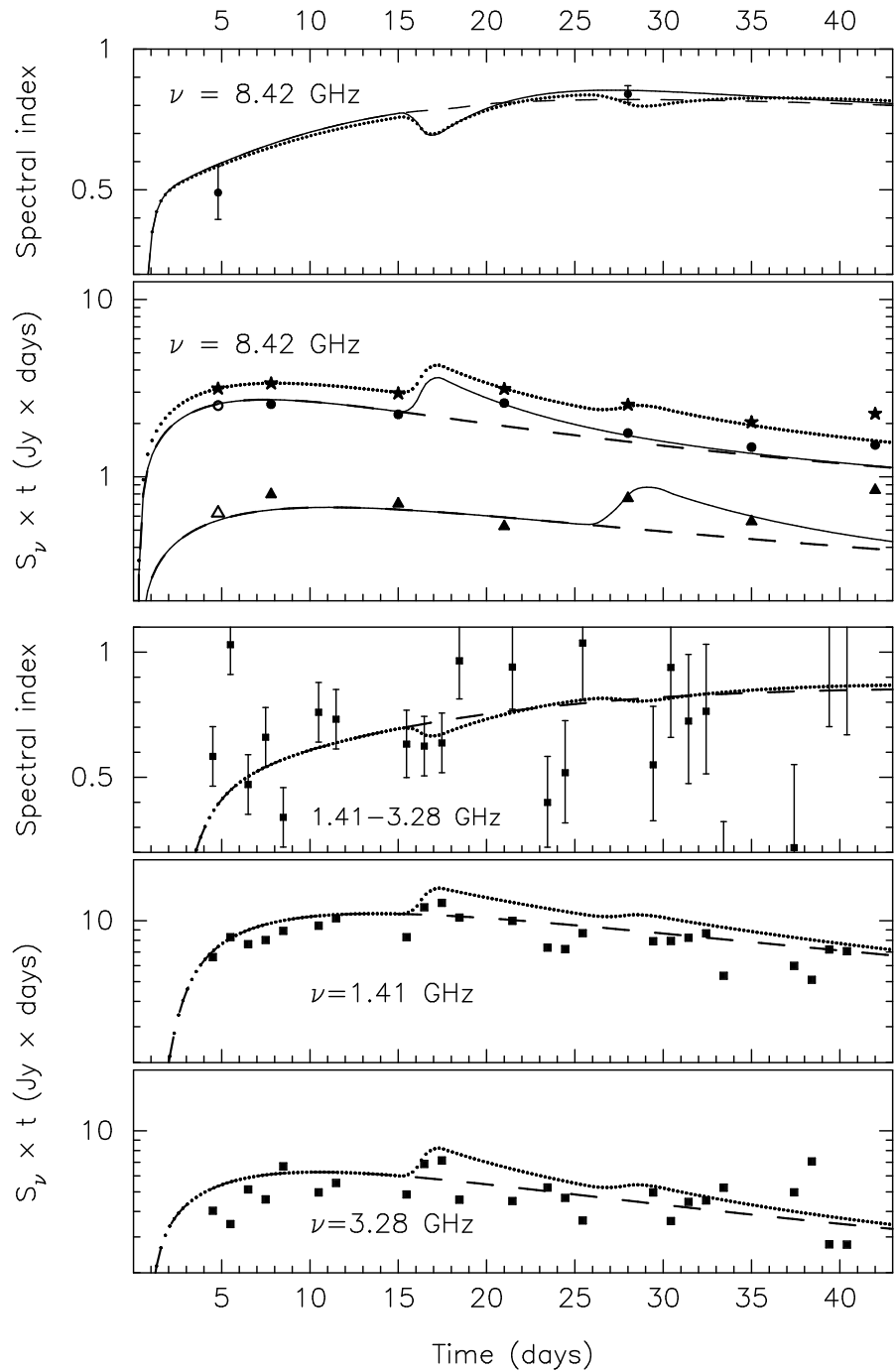}
\caption{The same as in Fig.8, including the Nancy data 
at 1.41\,GHz and 3.28\,GHz (Rodriguez et al. 1995). 
The dotted curves in this figure show the evolution of the 
total fluxes and relevant spectral indexes.}  
\end{figure}

\clearpage

%fig.10
\begin{figure}
\epsfxsize=13.5 cm
\epsffile[25 418 504 715]{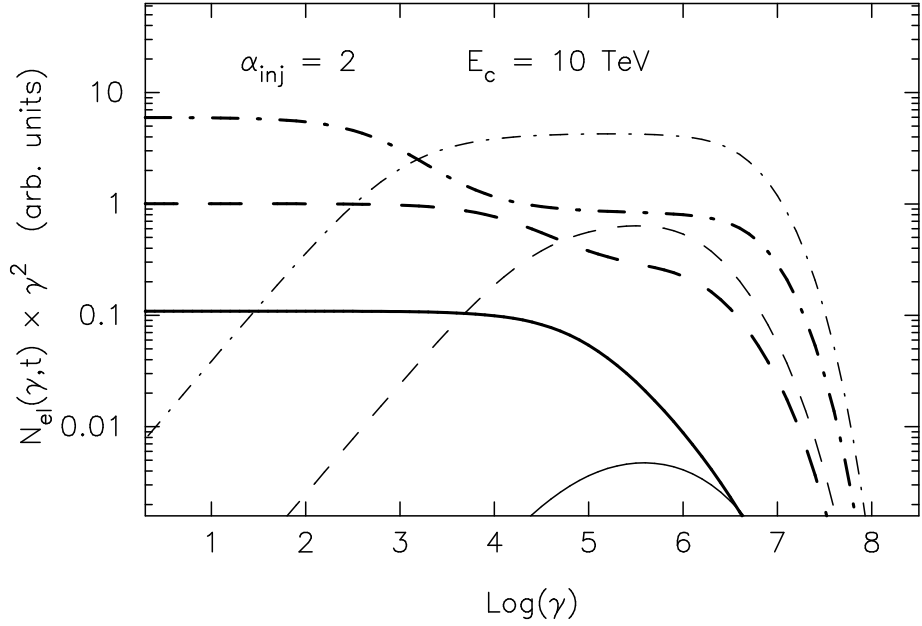}
\caption{Energy distribution of relativistic electrons inside (heavy lines)
and outside (thin lines) of the approaching radio cloud at the apparent times
$t = 0.1\,\rm day$ (solid lines),  1\,day (dashed lines), and 
10\,days (dot-dashed lines). Typical model parameters are supposed: $\alinj=2$,
$v_0= 0.2\,c$, $t_{\rm exp}=3\,\rm days$, $k=1$, $\Delta=1$, $u=1.5$, and
$C_\lambda =0.3$. The magnetic field 
at the instant $t_0 = 4.8\,\rm days$ is 
equal to $B_0 = 0.05\,\rm G$,
and the exponential cutoff energy for the injected electrons 
$E_{\rm c} = 10\,\rm TeV$.}
\end{figure}

\clearpage

%fig.11
\begin{figure}
\epsfxsize=14.5 cm
\epsffile[29 423 549 710]{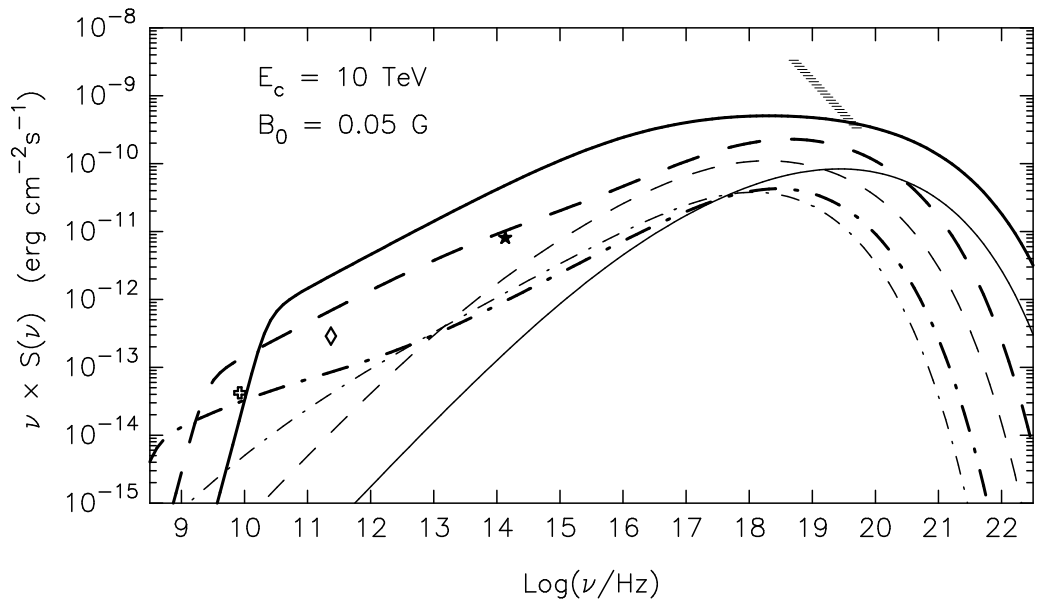}
\caption{The spectra of synchrotron radiation, produced inside (heavy lines) 
and outside (thin lines) of the clouds, by the
electrons shown in Fig.~10, at $t=0.1\,\rm day$ (solid lines), 
$t=1\,\rm day$ (dashed lines), and $t=10\,\rm days$ (dot--dashed lines). 
The cross and the diamond correspond to the level of maximum 
fluxes observed during the flares of GRS 1915+105 at $\nu = 8.42\,\rm GHz$ 
and 234\,GHz, respectively (Rodriguez et al. 1995). The star shows the flux 
of the IR jet (Sams et al. 1996), and the hatched region shows the level of
the hard X-ray fluxes typically detected from the source in the energy range 
$\geq 20\,\rm keV$ during the X-ray flares (e.g. Harmon et al. 1997).}
\end{figure}   

\clearpage

%fig.12
\begin{figure}
\epsfxsize=15.5 cm
\epsffile[46 423 559 671]{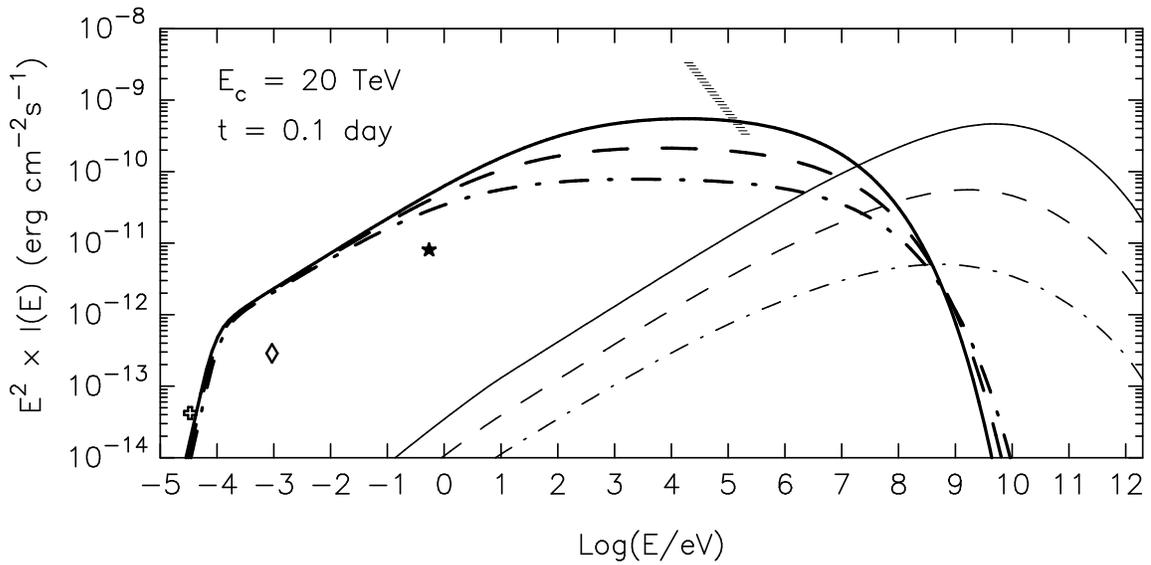}
\caption{The fluxes of the synchrotron (heavy lines) and IC (thin lines) 
radiations which could be 
expected from GRS 1915+105 at $t=0.1\,\rm day$ after 
ejection of a pair of radio clouds, calculated in the case of exponential 
cut-off energy $E_{\rm c}=20\,\rm TeV$, assuming 3 different values 
for the magnetic field at the instant $t=t_0$: $B_0=0.05\,\rm G$ (solid lines)
$B_0=0.1\,\rm G$ (dashed line), and $B_0=0.2\,\rm G$ (dot--dashed lines). 
All other model parameters are the same as in Fig.11.}
\end{figure}

\clearpage

%fig.13
\begin{figure}
\epsfxsize=13.5 cm
\epsffile[30 420 476 681]{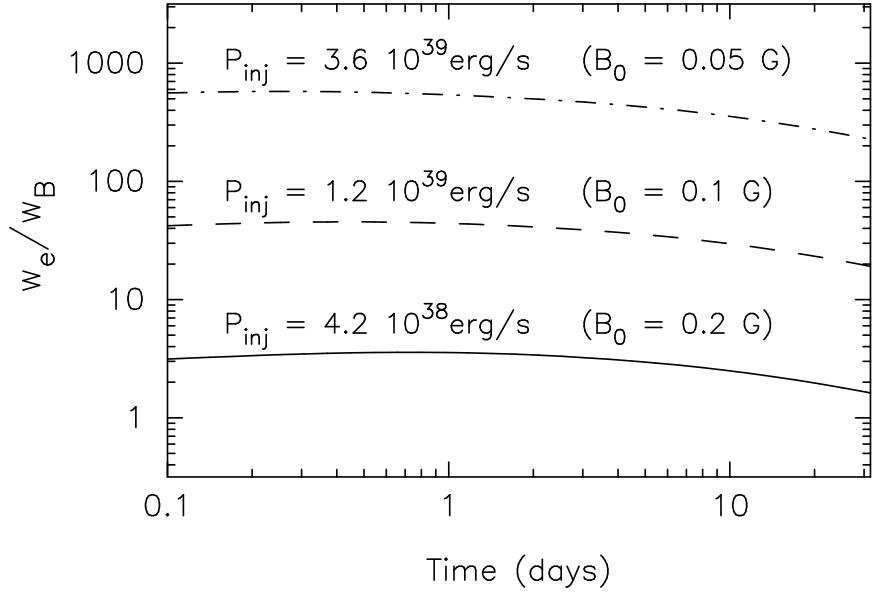}
\caption{The ratios of the electron to the magnetic field energy densities
in the expanding plasmoids corresponding to the calculations in Fig.12. 
Different values of the magnetic field $B_0$ at the instant 
$t_0$ (when the cloud's radius $R=R_0 \sim 10^{15}\,\rm cm$) 
actually imply different 
injection power of relativistic electrons $P_{\rm inj}$. For the supposed  
$E_{\rm c} = 20\,\rm TeV$,  the 
`equipartition' magnetic field  
corresponds to $B_0\simeq 0.3\,\rm G$.}
\end{figure}

\clearpage

%fig.14
\begin{figure}
\epsfxsize=15.5 cm
\epsffile[30 58 570 684]{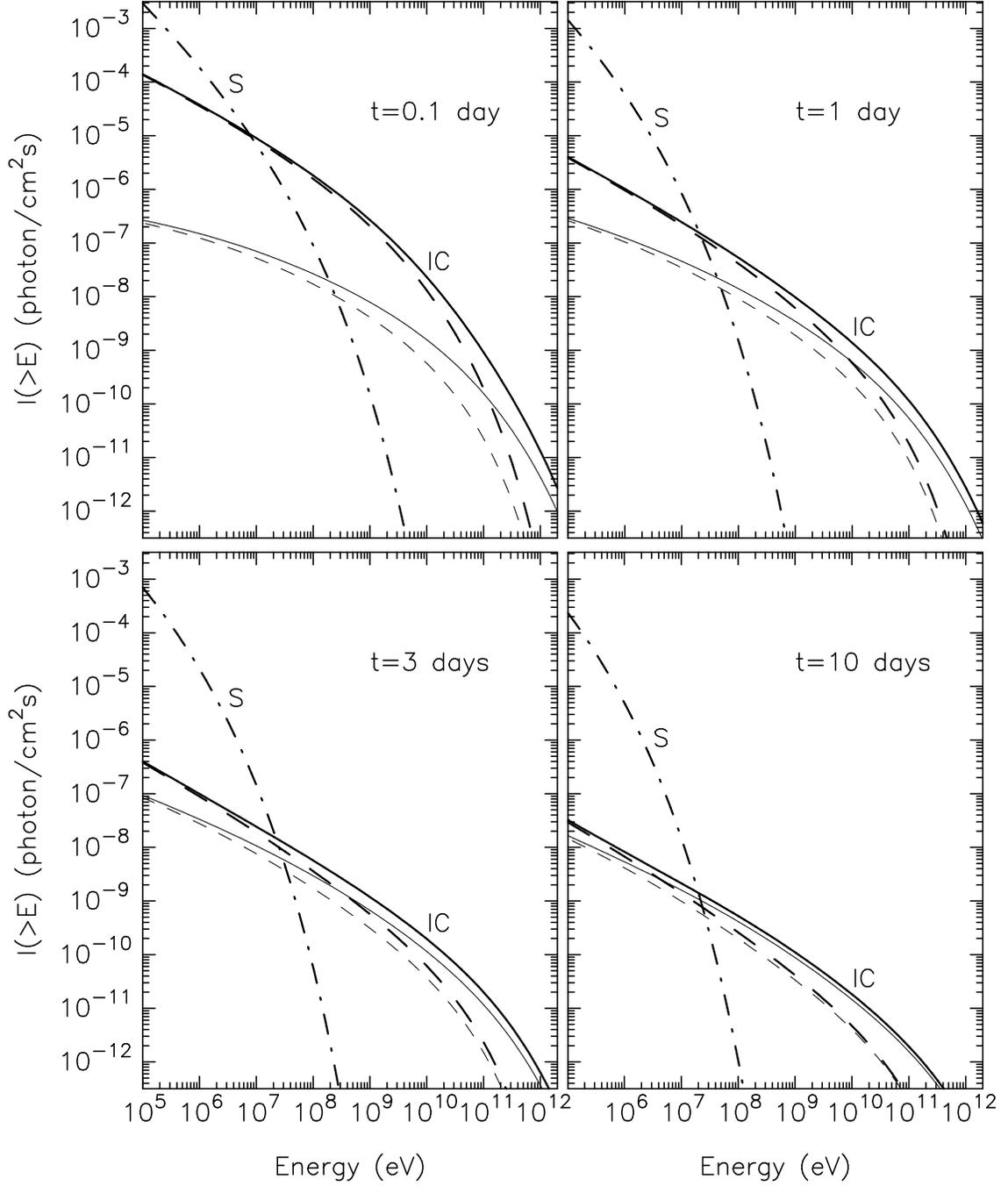}
\caption{The fluxes of IC $\gamma$-rays expected at different times 
$t$ from GRS 1915+105 
during a strong radio flare in the case of exponential cut-off energy
$E_{\rm c}= 20\,\rm TeV$ (solid lines) and $E_{\rm c}= 1\,\rm TeV$ 
(dashed lines). The fluxes produced inside and outside of the radio clouds
(see text) are plotted by heavy and thin lines, respectively. The dot--dashed
curves show the fluxes of synchrotron radiation for 
$E_{\rm c}= 20\,\rm TeV$.}
\end{figure}

\clearpage

%fig.15
\begin{figure}
\epsfxsize=12.5 cm
\epsffile[15 54 459 599]{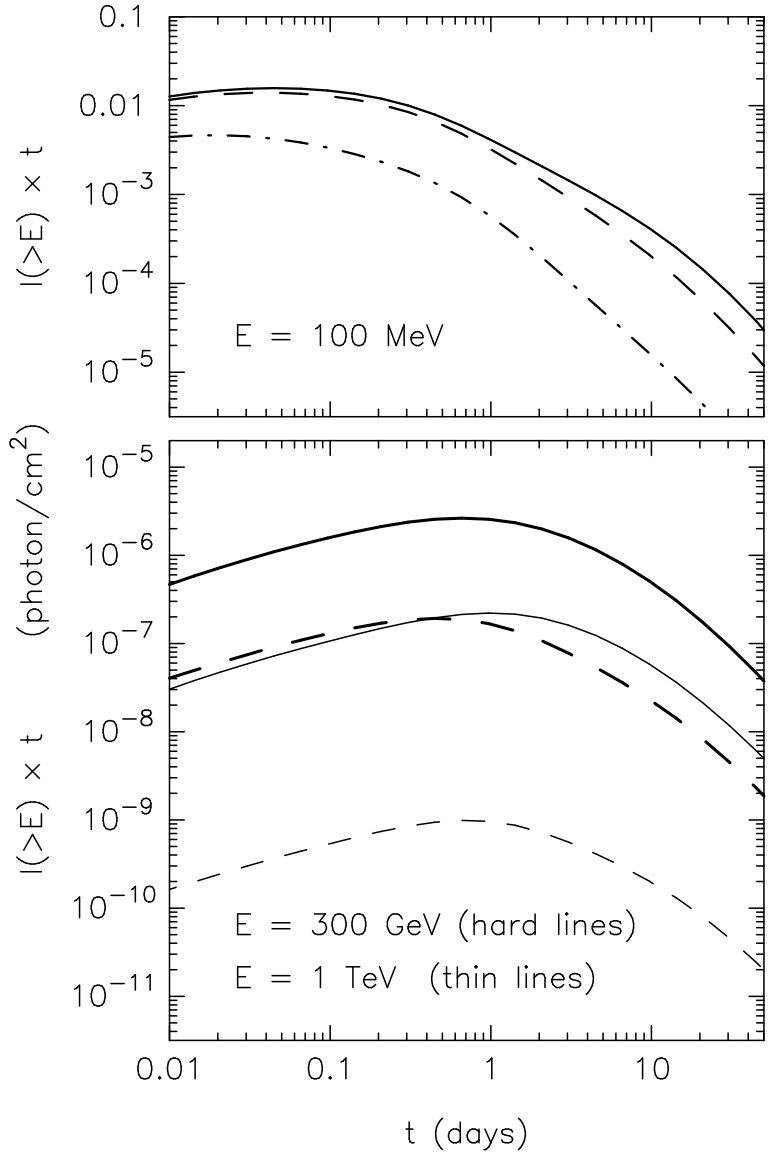}
\caption{The fluences of the high and very-high energy
IC $\gamma$-rays expected during strong radio flares in GRS 1915+105 
in the case of 3 different exponential cut-off energies for the accelerated 
electrons:
$E_{\rm c}= 20\,\rm TeV$ (solid curves), $E_{\rm c}= 1\,\rm TeV$ 
(dashed curves), and $E_{\rm c}= 30\,\rm GeV$ (dot--dashed curve). 
All other model parameters are the same as in Fig.~11. 
The supposed magnetic field $B_0=0.05\,\rm G$ implies an injection
of relativistic electrons with the initial (i.e. at $t\ll t_{\rm inj}$) power 
$P_{\rm inj} \simeq (3 \pm 0.6)\times 10^{39}\, \rm erg/s$, depending on 
$E_{\rm c}$.}
\end{figure}

\end{document}